\documentclass{article}

\usepackage{PRIMEarxiv}

\usepackage[utf8]{inputenc} 
\usepackage[T1]{fontenc}    
\usepackage{hyperref}       
\usepackage{url}            
\usepackage{booktabs}       
\usepackage{amsfonts}       
\usepackage{nicefrac}       
\usepackage{microtype}      
\usepackage{lipsum}
\usepackage{fancyhdr}       
\usepackage{graphicx}       
\graphicspath{{media/}}     

\usepackage{natbib}
\usepackage[shortlabels]{enumitem}
\usepackage{comment}
\usepackage{amsthm}
\usepackage{amsmath}
\usepackage{amssymb}
\usepackage{algorithm}
\usepackage{algpseudocode}
\usepackage{multirow}
\usepackage{array}

\theoremstyle{thmstyleone}%
\newtheorem{theorem}{Theorem}
\newtheorem{assumption}{Assumption}
\theoremstyle{thmstyletwo}%
\newtheorem{remark}{Remark}%
\theoremstyle{thmstylethree}%

\DeclareMathOperator*{\argmin}{arg\,min}

\DeclareMathOperator*\uplim{\overline{lim}}
\newcommand{\ind}{\perp\!\!\!\!\perp} 
\newcommand{\EYT}{g_1}
\newcommand{\EYC}{g_0}
\newcommand{\pSol}{\hat{\omega}^{p}}
\newcommand{\Ufun}{\overline{U}}
\newcommand{\phifun}{\overline{\phi}}
\newcommand{\Xlen}{d}
\newcommand{\norm}[1]{\left\lVert#1\right\rVert}

\newcolumntype{M}[1]{>{\centering\arraybackslash}m{#1}}

\title{Doubly robust causal inference through penalized bias-reduced estimation: combining non-probability samples with designed surveys}

\author{%
  Jiacong Du\\
  PhD student\\
  Department of Biostatistics\\
  University of Michigan\\
  Ann Arbor, MI, 48109 \\
  \texttt{jiacong@umich.edu} \\
  \And
  Xu Shi\thanks{Corresponding author}\\
  Assistant Professor \\
  Department of Biostatistics\\
  University of Michigan\\
  Ann Arbor, MI, 48109 \\
  \texttt{shixu@umich.edu} \\
  \AND
  Donglin Zeng \\
  Professor\\
  Department of Biostatistics\\
  University of Michigan\\
  Ann Arbor, MI, 48109 \\
  \texttt{dzeng@umich.edu} \\
  \And
  Bhramar Mukherjee \\
  Professor\\
  Department of Biostatistics\\
  University of Michigan\\
  Ann Arbor, MI, 48109 \\
  \texttt{bhramar@umich.edu} \\
}

\begin{document}
\maketitle

\begin{abstract}
Causal inference on the average treatment effect (ATE) using non-probability samples, such as electronic health records (EHR), faces challenges from sample selection bias and high-dimensional covariates. This requires considering a selection model alongside treatment and outcome models that are typical ingredients in causal inference. This paper considers integrating large non-probability samples with external probability samples from a design survey, addressing moderately high-dimensional confounders and variables that influence selection. In contrast to the two-step approach that separates variable selection and debiased estimation, we propose a one-step plug-in doubly robust (DR) estimator of the ATE. We construct a novel penalized estimating equation by minimizing the squared asymptotic bias of the DR estimator. Our approach facilitates ATE inference in high-dimensional settings by ignoring the variability in estimating nuisance parameters, which is not guaranteed in conventional likelihood approaches with non-differentiable L1-type penalties. We provide a consistent variance estimator for the DR estimator. Simulation studies demonstrate the double robustness of our estimator under misspecification of either the outcome model or the selection and treatment models, as well as the validity of statistical inference under penalized estimation. We apply our method to integrate EHR data from the Michigan Genomics Initiative with an external probability sample.
\end{abstract}


\section{Introduction}

Randomized clinical trials (RCTs) have long been considered the gold standard for assessing new treatments. However, many practical problems arise when collecting and analyzing RCT data, such as elevated costs, time constraints, and patient adherence \citep{karanatsios2020defining, herbert2018analysis}. Alternatively, the increasing availability of cost-efficient non-probability samples, such as electronic health records (EHR), presents new opportunities to study comparative effectiveness and safety in clinical research using causal inference methods \citep{mc2018routinely}. One significant advantage of EHR data is that it contains rich patient information and a far wider spectrum of disease phenotypes than many RCTs. As such, EHR data offer researchers the opportunity to explore the treatment-outcome relationship beyond the original RCT, advance pharmacovigilance drug repurposing, create synthesis controls, and achieve adaptive design \citep{suchard2019comprehensive,hripcsak2015observational, lauffenburger2021rationale, xu2015validating}.

However, the use of EHR data still faces methodological challenges that are often ignored in many analyses. As a non-probability sample, one concern is the presence of selection bias \citep{beesley2020emerging, beesley2022statistical, baker2013summary}. Since EHR data are primarily collected for clinical care and billing purposes, the selection mechanism of patients into the EHR system is unknown. As a result, the EHR dataset lacks the representativeness of a defined target population, e.g. the US population, which brings difficulty when generalizing findings beyond the sample.

To correct selection bias in EHR data, external probability samples, such as designed surveys, are useful since they are selected under a known sampling design and are thus representative of the target population. Three common estimators of the average treatment effect (ATE) are used for integrating a non-probability sample and another probability sample, including the outcome regression (OR) estimator, the inverse probability weighting (IPW) estimator, and the doubly robust (DR) estimator \citep{shi2022data, dahabreh2019generalizingWithSubsample, dahabreh2019generalizingToAll}. These estimators use observed treatment, outcome and covariates information in the internal non-probability sample as well as the sampling weights and covariates in the external probability sample, but do not need treatment or outcome information from the external probability sample. The OR estimator requires correctly specified outcome models for consistent estimation of the ATE while the IPW estimator requires correctly specified selection and treatment models. The DR estimator combines OR and IPW estimators and remains consistent if either the outcome models or the weighting models (i.e. both selection and treatment models) are correctly specified.

Despite the growing methodological development in combining probability and non-probability samples for causal inference, another challenge arises from the large number of potential confounders and variables influencing selection into the study. With an unknown selection mechanism and incomplete understanding of the confounding structure influencing the treatment-outcome relationship, researchers often need to explore a large number of variables from both datasets. This leads to complications in modeling the outcome, selection, and treatment mechanisms, all potentially with high-dimensional predictors. In this case, variable selection becomes necessary in estimating nuisance parameters in the three working models to address the instability and infeasibility of the maximum likelihood estimation. 

However, to the best of our knowledge, there is a lack of methods that consider a combination of the three statistical problems in causal inference, data integration and high-dimensional data. In this paper, we aim to make causal inference of the ATE by integrating a non probability sample with an external probability sample, where there is a large number of potential confounders and selection variables present in both datasets. The key challenge is to provide valid inference of the ATE given that variable selection is needed to handle high dimensional covariates in the outcome, treatment and selection models.

Existing literature has largely focused on addressing these challenges in pairs, typically tackling two of the three issues simultaneously. 
Specifically, there is a rich literature on adapting penalization techniques for causal inferences in the presence of high-dimensional confounders, where only the outcome and treatment models are considered \citep{belloni2014inference, farrell2015robust, chernozhukov2018double, chernozhukov2018generic}. Several methods have been developed in this field, including double/debiased machine learning \citep{chernozhukov2018double, chernozhukov2018generic} and the outcome adaptive LASSO \citep{shortreed2017outcome, ju2020robust, kabata2021variable}. These methods typically require correctly specified outcome and treatment models to make valid inferences. Based on the work of \cite{vermeulen2015bias}, \cite{avagyan2021high} proposed a LASSO-penalized bias-reduced method that produces robust inference under model misspecification. The robustness is achieved by using the estimating equation that minimizes the squared first-order asymptotic bias of the DR estimator. 
More recently, there is a growing literature on integrating probability and non-probability samples in high-dimensional settings. With the goal of estimating the population mean, current literature only considers the outcome and selection models. 
Researchers have proposed various methods to estimate the selection probability, such as model-based calibration using the LASSO penalty \citep{tsung2018model} and outcome adaptive LASSO \citep{bahamyirou2021data}. 
\cite{yang2020doubly} developed a two-step method for finite sample inference of population mean. The first step uses a penalized likelihood-based estimating equation procedure with the SCAD penalty to identify variables associated with the outcome and the selection. The second step re-estimates the nuisance parameters based on the selected set of variables from the first step, which facilitates the downstream calculation of the asymptotic distribution of the DR estimator of the population mean.

For our problem, a two-step method that separates variable selection and debiased estimation has been implied from the existing literature to handle high-dimensional covariates while achieving the doubly robust inference.
However, this approach places strict requirements on the variable selection procedure, requiring the identification of all true predictors \citep{yang2020doubly,cho2023variable}. 
In contrast, we propose a novel one-step, plug-in DR estimator of the ATE. When using the DR estimator, it is important to consider the impact of the estimators for the nuisance parameters in the three workings models on the asymptotic distribution of the DR estimator of the ATE. 
According to \cite{vermeulen2015bias}, when all working models are correctly specified, any root-n consistent estimators of the nuisance parameters will yield the same asymptotic distribution of the DR estimator of the ATE, as the first-order derivative of the DR estimator is asymptotically zero. Nevertheless, depending on the choice of the estimators of the nuisance parameters, this property might be lost when one of the working models is misspecified. In this case, it is necessary to account for the uncertainty of the nuisance parameter estimators in the first-order bias of the DR estimator when deriving its asymptotic distribution. However, this cannot be achieved if nuisance parameters are estimated from conventional penalized likelihood approaches with possibly misspecified models since the penalty function, e.g., LASSO penalty, is non-differentiable at zero. 

To address this issue, we propose a novel penalized estimating equation for simultaneous variable selection and parameter estimation for the selection, treatment, and outcome working models. Our penalized estimating equation minimizes the squared asymptotic bias of the DR estimator, such that the first-order derivative of the DR estimator with respect to the nuisance parameters is $o_p(1)$, as opposed to $O_p(1)$ from conventional penalized approaches. As a result, the uncertainty of the estimators of the nuisance parameters can be ignored asymptotically when making the inference for ATE, and thus it facilitates the calculation of the asymptotic distribution of the DR estimator. We focus on the penalty functions that achieve the ``oracle properties'': i) selection consistency, ii) producing a sparse solution, and iii) preserving large coefficients, such as the SCAD and the aLASSO penalties.

To summarize, we make the following contributions to the literature: Firstly, we establish the identification results of the ATE when integrating a non-probability sample with a probability sample. Secondly, given high-dimensional predictors, we extend the bias-reduced estimation approach proposed by \cite{vermeulen2015bias} to estimate nuisance parameters in the working models. Our extension is two-fold. The first extension generalizes the bias-reduced estimation procedure to account for three working models of the outcome, treatment, and selection. The second extension incorporates penalization into the new estimating equations to perform simultaneous variable selection and coefficient estimation. We show how our proposed penalized estimating equations can aid in making inferences of the ATE using the DR estimator under high-dimensional settings.

The rest of the paper is organized as follows: In Section 2, we present our proposed penalized estimating equation procedure for variable selection and estimation for nuisance parameters in the working models. Section 3 describes the computational algorithm for solving the penalized estimating equation using a local quadratic approximation for the penalty function and an iterative Newton-Raphson updates. To enhance computational efficiency and to avoid large matrices inverse, our algorithm separates the optimization for the outcome models and the weighting models. Section 4 presents the theoretical properties of the penalized estimating equation procedure and the asymptotic distribution and variance estimation for the DR estimator. Simulation studies to evaluate the performance of our proposed strategy are shown in Section 5. In Section 6, we apply our method to integrate EHR data from the Michigan Genomics Initiative (MGI), a longitudinal biorepository at the University of Michigan, with a probability sample from the National Health and Nutritional Examination Survey (NHANES) to estimate the causal effect of severe obesity on systolic blood pressure/hypertension. We conclude with a brief discussion in Section 7.

\section{Proposed method}

Denote $Y$ as the outcome, $T$ as the treatment indicator, and $X$ as a vector of covariates of length $\Xlen$ (including the intercept). Consider a probability sample $A$ of size $n_A$ and a non-probability sample $B$ of size $n_B$ drawn from the target population of size $N$. We assume that there is no overlap between sample $A$ and $B$. Let $I_{A,i}$ be the indicator for selecting person $i$ into sample $A$ and $I_{B,i}$ be the indicator for selecting person $i$ into sample $B$. Let $d_{A,i}$ denote the sampling weight for the $i$-th person, that is, $d_{A,i} = 1/P(I_{A,i}=1|X=x_i)$. \textbf{Table \ref{data_source}} illustrates the structure of the two datasets and the target population. Data on covariates and sampling weights are available in sample $A$, while data on covariates, outcome, and treatment are available in sample $B$. That is, we have $\{(x_i,d_{A,i}), i\in A \}$ in sample $A$ and $\{(x_i,y_i, t_{i}), i\in B \}$ in sample $B$. 
In practice, while confounders and selection variables are likely to be different, it is common to start with an inclusive large set of variables due to unknown selection mechanism and incomplete understanding of the confounding structure influencing treatment-outcome relationship. Thus, the covariate vector $X$ contains candiate predictors for both selection mechanism and the outcome. 
We define a pair of potential outcomes $\big(Y(1), Y(0) \big)$, that would be observed had the person been given treatment, $Y(1)$, or control, $Y(0)$. The outcome type could be continuous or binary. The parameter of interest is the ATE, denoted as $\theta = E\{Y(1)-Y(0)\}$.

\begin{table}[H]
\centering
\caption{Illustration of the structure of the two data sources and the target population. ``$\checkmark$'' and ``?'' indicate observed and possibly unobserved data, respectively.} \label{data_source}
\begin{tabular}{p{3cm}ccccc} 
\hline 
Sample                 & id        & \begin{tabular}[c]{@{}c@{}}Sampling\\ weight\end{tabular} & \begin{tabular}[c]{@{}c@{}}Covariates\\ $X$\end{tabular} & \begin{tabular}[c]{@{}c@{}}Treatment\\ $T$\end{tabular} & \begin{tabular}[c]{@{}c@{}}Outcome\\ $Y$\end{tabular} \\ 

\multirow{3}{*}{\begin{tabular}[c]{@{}l@{}}Probability sample $A$\\ $I_A=1, I_B=0$\end{tabular}} & 1         &        \checkmark         &     \checkmark         &     ?        &       ?    \\
                       &    \vdots       &       \vdots          &     \vdots         &     \vdots        &    \vdots       \\
                       & $n_A$      &       \checkmark          &     \checkmark         &       ?      &      ?     \\ 
\multirow{3}{*}{\begin{tabular}[c]{@{}l@{}}Non-probability sample $B$\\ $I_A=0, I_B=1$\end{tabular}} & $n_A+1$    &         ?        &     \checkmark         &      \checkmark       &     \checkmark      \\ 
                       &    \vdots       &       \vdots          &     \vdots         &    \vdots         &      \vdots     \\
                       & $n_A+n_B$ &       ?          &       \checkmark       &    \checkmark         &     \checkmark     \\ 
\multirow{3}{*}{\begin{tabular}[c]{@{}l@{}}Other individuals in \\ the target population\\ $I_A=I_B=0$\end{tabular}} & $n_A+n_B+1$    &         ?        &     ?         &      ?       &     ?      \\ 
&    \vdots       &       \vdots          &     \vdots         &    \vdots         &      \vdots     \\
                       & $N$ &       ?          &
                       ?       &    ?      
                       &     ?     \\ \hline
\end{tabular}
\end{table}

\subsection{Identification}

The following assumptions are commonly imposed for identifying the ATE with integrated data \citep{shi2022data,dahabreh2019generalizingToAll,dahabreh2019generalizingWithSubsample,dahabreh2019efficient}:

\begin{assumption}\label{assumption1}
We assume 
\begin{enumerate}[label=(\alph*)]
\item \label{A1_a}%
    Consistency: $Y=Y(t)$ if $T=t$, $t=0,1$;
\item \label{A1_b}%
    Treatment exchangeability: $Y(t)\ind T|X,I_B=1$; 
\item\label{A1_c}%
    Treatment positivity: $P(T=t|X,I_B=1)>0$; 
\item \label{A1_d}%
    Selection exchangeability: $Y(t)\ind I_B|X$;  
\item \label{A1_e}%
    Selection positivity: $P(I_B=s|X)>0$, where $s=0,1$.
\end{enumerate}
\end{assumption}
Assumptions \ref{assumption1}\ref{A1_a} - \ref{assumption1}\ref{A1_c} allow us to generalize the mean model of the observed outcome to that of the potentially unobserved potential outcome within sample $B$: $E(Y|X,T=t,I_B=1) =  E\{Y(t)|X,I_B=1\}$, while Assumptions \ref{assumption1}\ref{A1_d} and \ref{assumption1}\ref{A1_e} allow us to generalize the mean model of the potential outcome from sample $B$ to the target population: $E\{Y(t)|X,I_B=1\} = E\{Y(t)|X\}$.

Under the above identifiability conditions, $\theta$ is identifiable via the g-formula \citep{shi2022data,dahabreh2019generalizingWithSubsample,dahabreh2021study} by
\begin{equation*}
    \theta = E\big[ I_A d_A \big\{ E(Y|X,I_B=1,T=1) - E(Y|X,I_B=1,T=0) \big\} \big].
\end{equation*}
\noindent 
The outer expectation is taken with respect to the joint distribution of $I_A$ and $X$ in the target population. This approach models the conditional expectation of the outcome given the treatment or the control and is valid if both models are correctly specified. On the other hand, $\theta$ is also identifiable via weighting \citep{shi2022data,dahabreh2021study} by 
\begin{equation*}
    \theta = E\bigg\{\frac{I_B T}{P(I_B=1, T=1|X)}Y -  \frac{I_B (1-T)}{P(I_B=1, T=0|X)}Y \bigg\}.
\end{equation*}
The weighting approach models $P(I_B=1, T=t|X)$ as the weights (inverse) for the observed outcome for each individual in sample $B$. Although one can directly model the joint distribution of $I_B$ and $T$, this approach is less commonly seen due to the complexity of capturing both the selection and treatment mechanisms in one model. On the other hand, 
a common practice involves with decomposing the joint probability as the product of sequential conditional probabilities, $P(I_B=1, T=t|X)=P(I_B=1|X)\cdot P(T=t|I_B=1,X)$, and estimate the selection probability $P(I_B=1|X)$ and the treatment probability $P(T=t|I_B=1,X)$ separately. Another decomposition of $P(I_B=1, T=t|X)$ is $P(I_B=1|X,T=t) \cdot P(T=t|X)$. However, estimating these two probabilities would need the treatment information in sample $A$, which might be unavailable. Therefore, we do not consider the last type of decomposition in this paper. 

In general, the weighting approach is valid if $P(I_B=1, T=t|X)$ is correctly modeled. In this paper, we will mainly describe our method based on the decomposition of $P(I_B=1, T=t|X)=P(I_B=1|X)\cdot P(T=t|I_B=1,X)$. Nevertheless, our method can be easily extended to the direct modeling of $P(I_B=1, T=t|X)$ and we will perform simulation studies to show that. 

\subsection{Existing estimators}

The estimation of $\theta$ relies on estimating $E(Y|X,I_B=1,T=t)$ and/or $P(I_B=1|X), P(T=t|I_B=1,X)$. 
Since non-parametric estimation of these components may not be feasible if $X$ is high-dimensional, dimension-reducing parametric models are often used for estimation. As such, we postulate the following working models, which may be misspecified:
\begin{enumerate}[label={(\alph*)}] 
\item \label{A2_a}
    Selection model: $P(I_B=1|X) = \pi_B(X;\alpha)$;
\item \label{A2_b}
    Treatment model: $P(T=1|X,I_B=1)=\pi_T(X;\tau)$;
\item\label{A2_c}
    Expected outcome given the treatment in sample $B$: $E(Y|X, T=1, I_B=1) = \EYT (X;\beta)$;
\item \label{A2_d}
    Expected outcome given the control in sample $B$: $E(Y|X, T=0, I_B=1) = \EYC(X;\gamma)$,
\end{enumerate}
\noindent where $\alpha, \tau, \beta, \gamma \in \mathbb{R}^p$ are unknown nuisance parameters, with true values of $\alpha_0$, $\tau_0$, $\beta_0$ and $\gamma_0$. Let $\theta_0$ denote the true value of the ATE. Implied by the above identification results, we present three types of estimators for ATE in the existing literature \citep{shi2022data,dahabreh2019generalizingToAll,yang2020doubly}: the OR estimator, which relies on modeling the mean of the outcome; the IPW estimator, which relies on modeling the selection probability and the treatment probability; and the DR estimator, which combines the OR and IPW estimators. We assume that the target population size $N$ is known. If not, $N$ is usually estimated using the sampling weights from sample $A$ \citep{sarndal2003model}.

The OR estimator is given by: 
\begin{equation}\label{OR_estimator}
    \hat{\theta}_{\rm OR} = N^{-1} \sum_{i=1}^N I_{A,i}d_{A,i} \big\{\EYT(X_i;\hat{\beta}) - \EYC(X_i;\hat{\gamma}) \big\}.
\end{equation}
Typically, $\hat{\beta}$ and $\hat{\gamma}$ can be obtained by fitting a regression model for the observed outcome in the treated and control group in sample $B$. When the outcome models are correctly specified, $\hat{\beta} \xrightarrow{p} \beta_0$ and $\EYT(X;\hat{\beta}) \xrightarrow{p} E(Y|X,I_B=1,T=1)$, where ``$\xrightarrow{p}$'' denotes ``converge in probability''. Similarly, $\EYC(X;\hat{\gamma}) \xrightarrow{p} E(Y|X,I_B=1,T=0)$. Then we have a consistent estimator for $\theta$, that is, $\hat{\theta}_{\rm OR} \xrightarrow{p} \theta_0$. The OR estimator requires correctly specified outcome models in each treatment group for a consistent estimator of the ATE. 

The IPW estimator is described as:
\begin{equation}\label{IPW_estimator}
    \hat{\theta}_{\rm IPW} = N^{-1} \sum_{i=1}^N \Big\{ \frac{I_{B,i}T_{i}}{\pi_{B}(X_i;\hat{\alpha}) \pi_{T}(X_i;\hat{\tau})} Y_i - \frac{I_{B,i}(1-T_{i} )}{\pi_{B}(X_i;\hat{\alpha}) \big(1-\pi_{T}(X_i;\hat{\tau})\big)} Y_i \Big\}.
\end{equation}
To estimate $\tau$, people often fit a regression model for the treatment indicator using sample $B$. In contrast, $\alpha$ is often estimated based on the combined samples $A$ and $B$ via approaches such as calibration \citep{kott1990estimating} and maximum pseudo-likelihood estimation \citep{chen2020doubly}. The IPW estimator is consistent if $\pi_B(X;\alpha)$ and $\pi_T(X;\tau)$ 
are correctly specified. 

The DR estimator combines the previous two estimators to gain robustness under model misspecification: \citep{shi2022data,dahabreh2019generalizingWithSubsample,dahabreh2019efficient,yang2020doubly}:
\begin{align}\label{DR_estimator_existing}
    \hat{\theta}_{\rm DR} = \frac{1}{N} \sum_{i=1}^N \Big[ 
    &\Big\{ I_{A,i} d_{A,i} \EYT(X_i;\hat{\beta}) + \frac{I_{B,i} T_i}{\pi_B(X_i;\hat{\alpha}) \pi_T(X_i;\hat{\tau})}\big(Y_i-\EYT(X_i;\hat{\beta}) \big) \Big\} - \nonumber\\
    & \Big\{ I_{A,i} d_{A,i} \EYC(X_i;\hat{\gamma}) + \frac{I_{B,i} (1- T_i)}{\pi_B(X_i;\hat{\alpha}) \big(1-\pi_T(X_i;\hat{\tau})\big)} \big(Y_i-\EYC(X_i;\hat{\gamma}) \big) \Big\} \Big]. 
\end{align}
$\hat{\theta}_{\rm DR}$ is doubly robust in that the estimator is consistent if the weighting models, $\pi_{B}(X;\alpha), \pi_T(X;\tau)$, are correctly specified, or if the outcome models, $\EYT(X;\beta),\EYC(X;\gamma)$, are correctly specified, but not necessarily both. For this nice property, we will focus on the DR estimator in this paper.

\begin{remark}
\normalfont
Regarding the consistency of the IPW and DR estimators, it is possible that even though models for $P(I_B=1|X)$ and $P(T=t|X,I_B=1)$ are both misspecified under $\pi_B(X;\alpha)$ and $\pi_T(X;\tau)$, estimates for their probability are still consistent; that is, $\pi_B(X;\hat{\alpha}) \xrightarrow{p} P(I_B=1|X)$ and $\pi_T(X;\hat{\tau}) \xrightarrow{p} P(T=1|X,I_B=1)$. Or more general, estimates for the joint probability are still consistent; that is, $\pi_B(X;\hat{\alpha})\pi_T(X;\hat{\tau}) \xrightarrow{p} P(I_B=1,T=1|X)$ and $\pi_B(X;\hat{\alpha})\big(1-\pi_T(X;\hat{\tau})\big) \xrightarrow{p} P(I_B=1,T=0|X)$. In this case, the IPW and DR estimators would also be consistent. 
\end{remark}

\begin{remark}
\normalfont
An alternative approach is to directly model the joint distribution of $P(I_B=1,T=t|X)$ with the working models $w_1(X;\delta_1)$ for $P(I_B=1,T=1|X)$ and $w_0(X;\delta_0)$ for $P(I_B=1,T=0|X)$. The corresponding DR estimator of the ATE is 
\begin{align}
    \hat{\theta}_{\rm DR} = N^{-1} \sum_{i=1}^N \Big[
    & \Big\{ I_{A,i} d_{A,i} \EYT(X_i;\hat{\beta}) + \frac{I_{B,i} T_i}{w_1(X_i;\hat{\delta}_1)}\big(Y_i-\EYT(X_i;\hat{\beta}) \big) \Big\} - \nonumber\\
    & \Big\{ I_{A,i} d_{A,i} \EYC(X_i;\hat{\gamma}) + \frac{I_{B,i} (1- T_i)}{w_0(X_i;\hat{\delta}_0)} \big(Y_i-\EYC(X_i;\hat{\gamma}) \big) \Big\} \Big]. 
    \label{DRestimator2}
\end{align}
\end{remark}

\subsection{Estimating nuisance parameters}

Let $\omega = (\alpha^{\top}, \tau^{\top}, \beta^{\top}, \gamma^{\top})^{\top}$ contain all the nuisance parameters in the working models. In practice, $\omega$ is unknown and needs to be estimated from data. The conventional approach is to estimate $\alpha,\tau,\beta,\gamma$ from separate likelihood models. 
While $\hat{\theta}_{\rm DR}$ is consistent if either the weighting models or the outcome models are correctly specified, its asymptotic distribution is impacted under model misspecification since the first-order asymptotic bias of the DR estimator is non-zero. In this case, it is necessary to account for the uncertainty of the nuisance parameter estimators when deriving the asymptotic distribution of the DR estimator, namely, parameter estimators in the outcome, selection and treatment models. However, when $X$ is high-dimensional and the penalization technique is used for variable selection, this cannot be achieved from the penalized likelihood approach since the penalty function is non-differentiable at zero. 

To address this issue, we propose to estimate $\omega$ in a way such that the squared asymptotic bias of the DR estimator is minimized. 
While existing literature has explored bias-reduced estimation approaches, they have primarily focused on the outcome and treatment models \citep{vermeulen2015bias,avagyan2021high}. However, the incorporation of a selection model into this framework remains largely unexplored. Furthermore, little attention has been paid to explore the advantages of this approach in high-dimensional settings. 
Therefore, we first present the estimating method in low-dimensional settings and will extend it to high-dimensional settings in the next section.

Define
\begin{align*}
    \phi(Z;\theta,\omega) &=  \Big\{ I_{A} d_{A} \EYT(X;\beta) + \frac{I_{B} T}{\pi_B(X;\alpha) \pi_T(X;\tau)}\big(Y-\EYT(X;\beta) \big) \Big\} \\
    & - \Big\{ I_{A} d_{A} \EYC(X;\gamma) + \frac{I_{B} (1- T)}{\pi_B(X;\alpha) \big(1-\pi_T(X;\tau)\big)} \big(Y-\EYC(X;\gamma) \big) \Big\} - \theta,
\end{align*}
where $Z$ contains observed data in $(X,Y,T,I_A, d_A, I_B)$. The asymptotic bias of the DR estimator given $\theta_0$ is
\begin{equation*}
    bias(\omega; \theta_0) = E\{\phi(Z;\theta_0,\omega)\}.
\end{equation*}
\noindent To minimize the squared asymptotic bias, which is $bias^2(\omega; \theta_0)$, we note that $bias(\omega; \theta_0)$, is 0 if the weighting models, $\pi_B(X;\alpha)$ and $\pi_T(X;\tau)$, are correctly specified; or if the outcome models, $\EYT(X;\beta)$ and $\EYC(X;\gamma)$, are correctly specified. When all models are misspecified, $bias(\omega; \theta_0)$ is not necessarily zero \citep{vermeulen2015bias}. In this case, suppose there exists $\mathring{\omega}$ such that 
$\left. E\bigg\{\frac{\partial \phi(Z;\theta_0,\omega)}{\partial \omega^{\top}} \right \vert_{\omega=\mathring{\omega}} \bigg\} = 0$. Under regularity conditions in Section S1.1 in the supplementary materials, the squared asymptotic bias is minimized at $\mathring{\omega}$ since
\begin{equation*}
    \left. \frac{\partial bias^2(\omega; \theta_0)}{\partial \omega^{\top}} \right \vert_{\omega=\mathring{\omega}} = 2bias(\mathring{\omega}; \theta_0)\cdot 
    \left. E\bigg\{\frac{\partial \phi(Z;\theta_0,\omega)}{\partial \omega^{\top}} \right \vert_{\omega=\mathring{\omega}} \bigg\} = 0.
\end{equation*}
Therefore, we define the following empirical estimating function for $\omega$ based on $E\big\{ \partial_{\omega} \phi(Z;\theta_0,\omega) \big\}$, i.e., $\hat{\omega}$ solves $\Ufun(\omega)=0$, where 
\begin{align*}
    \Ufun(\omega) & = \partial_{\omega} \phifun(\omega) = 
    \begin{pmatrix}
 \partial_{\beta} \phifun(\omega) \\
 \partial_{\gamma} \phifun(\omega) \\
 \partial_{\alpha} \phifun(\omega) \\
 \partial_{\tau} \phifun(\omega)
\end{pmatrix} = N^{-1}\sum_{i=1}^N
\begin{pmatrix}
 \partial_{\beta} \phi(z_i; \omega) \\
 \partial_{\gamma} \phi(z_i; \omega) \\
 \partial_{\alpha} \phi(z_i; \omega) \\
 \partial_{\tau} \phi(z_i; \omega)
\end{pmatrix}.
\end{align*}
For simplicity, we denote $\eta = (\alpha^{\top}, \tau^{\top})^{\top}$ containing nuisance parameters in the weighting models, and $\mu= (\beta^{\top}, \gamma^{\top})^{\top}$ containing nuisance parameters in the outcome models. Let $\hat{\eta}$ and $\hat{\mu}$ be the solution obtained from solving $\Ufun(\omega)=0$. Further, let $\eta^\ast$ and $\mu^\ast$ be the converging values of $\hat{\eta}$ and $\hat{\mu}$, respectively.
Let $\eta_0$ and $\mu_0$ be the unknown true values of the model parameters if the postulated working models are correctly specified. In practice, it is likely that not all four models are correctly specified. The next theorem shows the conditions for obtaining consistent estimators from our proposed estimating equation while allowing model misspecification.

\begin{theorem} \label{consist_ATE}
Under suitable regularity conditions (see Section S1.1 in the supplementary materials), $\hat{\eta} \xrightarrow{p} \eta_0$, when the selection model $\pi_{B}(X;\alpha)$ and the treatment model $\pi_T(X;\tau)$ are correctly specified regardless of whether the outcome models are correctly specified or not. Similarly, $\hat{\mu} \xrightarrow{p} \mu_0 $, when the outcome models, $\EYT(X;\beta)$ and $\EYC(X; \gamma)$ are correctly specified regardless of whether the selection and treatment models are correctly specified or not. 
\end{theorem}

The proof of Theorem \ref{consist_ATE} is provided in Section S1.1 of the supplementary materials. Theorem \ref{consist_ATE} implies that if weighting models are correctly specified and outcome models are misspecified, the estimator for the selection and treatment probability in sample $B$ is still consistent for the truth, that is,
\begin{align*}
            \pi_T(X; \hat{\tau}) \xrightarrow{p} \pi_T(X; \tau_0) = P(T=1|X,I_B=1) & ,   \pi_B(X; \hat{\alpha}) \xrightarrow{p} \pi_B(X; \alpha_0) = P(I_B=1|X),
\end{align*}
\noindent while the estimator for the mean outcome may be inconsistent to the truth, that is,  
\begin{align*}
        \EYT(X; \hat{\beta}) \xrightarrow{p} \EYT(X; \beta^\ast) \neq E(Y|X,T=1,I_B=1) & ,\\
        \EYC(X; \hat{\gamma}) \xrightarrow{p} \EYC(X; \gamma^\ast) \neq E(Y|X,T=0,I_B=1).
\end{align*}
Similar arguments apply for $\hat{\mu}$. The property in Theorem \ref{consist_ATE} guarantees the doubly robustness of the DR estimator of the ATE, which we will discuss in more detail in Section 4. 

Theorem \ref{consist_ATE} also implies an iterative algorithm for solving $\omega$. Since $\hat{\eta} \xrightarrow{p} \eta_0$ when weighting models are correctly specified, we can fix $\mu$ as constant, i.e., $\Tilde{\mu}$, and only use 
$\begin{pmatrix}
\partial_{\beta} \phifun(\eta, \Tilde{\mu})\\
\partial_{\gamma} \phifun(\eta, \Tilde{\mu})
\end{pmatrix}$ to solve for $\eta$. Likewise, we can fix $\eta$ as $\Tilde{\eta}$ and use 
$\begin{pmatrix}
\partial_{\alpha} \phifun(\Tilde{\eta}, \mu)\\
\partial_{\tau} \phifun(\Tilde{\eta}, \mu)
\end{pmatrix}$ to solve for $\mu$. We will use this iterative method in the Computation section.

One limitation of this approach is that it requires the dimension of $\partial_{\beta} \phifun(\omega), \partial_{\gamma} \phifun(\omega), \partial_{\alpha} \phifun(\omega)$ and $\partial_{\tau} \phifun(\omega)$ in $\Ufun(\omega)$ to be the same \citep{vermeulen2015bias,yang2020doubly,chen2020doubly}, i.e., $|\partial_{\beta} \phifun(\omega)|=|\partial_{\gamma} \phifun(\omega)|=|\partial_{\alpha} \phifun(\omega)|=|\partial_{\tau} \phifun(\omega)|$, where $|\cdot|$ denotes the dimension of the vector function. Otherwise, the solution may not exist. Given the same set of $X$, this condition can be often satisfied if we postulate an additive model, i.e., $\pi_T(X;\tau) = \pi_T(X^\top\tau)$, $\pi_B(X;\alpha) = \pi_B(X^\top\alpha)$, $\EYT(X;\beta) = \EYT(X^\top\beta)$, and $\EYC(X;\gamma) = \EYC(X^\top\gamma)$. With some data transformation, interactions between covariates or pre-specified functions of a covariate, such as splines, can be handled by the additive model. For example, if we suspect that some working models should include higher-order terms, we can include such terms in all four working models so that the dimensions of $X$ are matched. 
Under the additive working models, $\Ufun(\omega)$ has a specific form that satisfies the above-mentioned dimension-matching criteria:
{\small
\begin{equation*}
    \Ufun(\omega)=
    \frac{1}{N}\sum_{i=1}^N \\
    \begin{pmatrix}
    \begin{aligned}
        I_{A,i}d_{A,i} \EYT^{(1)}(X_i^{\top}\beta)X_i - \frac{I_{B,i} T_i}{\pi_B(X_i^{\top}\alpha) \pi_T(X_i^{\top}\tau)}\EYT^{(1)}(X_i^{\top}\beta)X_i 
  \end{aligned}\\
    \begin{aligned}
        -I_{A,i}d_{A,i} \EYC^{(1)}(X_i^{\top}\gamma)X_i + \frac{I_{B,i} (1-T_i)}{\pi_B(X_i^{\top}\alpha) \big(1-\pi_T(X_i^{\top}\tau)\big)}\EYC^{(1)}(X_i^{\top}\gamma)X_i
    \end{aligned} \\
    \begin{aligned}
        I_{B,i}\bigg\{-\frac{T_i \big(Y_i-\EYT(X_i^{\top}\beta) \big)}{\pi_{T}(X_i^{\top}\tau)} + \frac{(1-T_i) \big(Y_i-\EYC(X_i^{\top}\gamma)\big)}{1-\pi_{T}(X_i^{\top}\tau)} \bigg\} \frac{\pi_B^{(1)}(X_i^{\top}\alpha)}{\big(\pi_B(X_i^{\top}\alpha)\big)^2} X_i
    \end{aligned} \\
    \begin{aligned}
        I_{B,i}\bigg\{ -\frac{ T_i \big(Y_i-\EYT(X_i^{\top}\beta)\big)}{\big( \pi_{T}(X_i^{\top}\tau) \big)^2 } - \frac{ (1-T_i) \big(Y_i-\EYC(X_i^{\top}\gamma) \big)}{\big( 1-\pi_T(X_i^{\top}\tau) \big)^2} \bigg\} \frac{\pi_{T}^{(1)}(X_i^{\top}\tau)}{\pi_{B}(X_i^{\top}\alpha)}  X_i
    \end{aligned}
    \end{pmatrix}, 
\end{equation*}
}

\noindent where $\EYT^{(1)}(X_i^{\top}\beta)$, $\EYC^{(1)}(X_i^{\top}\gamma), \pi_B^{(1)}(X_i^{\top}\alpha), \pi_T^{(1)}(X_i^{\top}\tau)$ denote the first-order derivative with respect to the linear predictor, $X_i^\top\beta, X_i^\top\gamma, X_i^\top\alpha, X_i^\top\tau$, respectively.

\subsection{Extension to high-dimensional settings}

So far, we have discussed the estimation equations for nuisance parameters in low-dimensional settings. We now consider the situation when $X$ is high-dimensional. In this case, we use the penalized estimating function to perform simultaneous variable selection 
and coefficient estimation \citep{fu2003penalized,johnson2008penalized}, given by  
\begin{equation} \label{Up_eq}
    \Ufun^p(\omega) = 
    \Ufun(\omega) + 
    \begin{pmatrix}
    q_{\lambda_\eta}(|\eta|)sgn(\eta) \\
    q_{\lambda_\mu}(|\mu|)sgn(\mu)
    \end{pmatrix},
\end{equation}
\noindent where $q_{\lambda}(u)=d p_{\lambda}(u)/du$ for a continuous penalty function  $p_{\lambda}(u)$ and $q_{\lambda}(|u|) = \big(q_{\lambda}(|u_0|), q_{\lambda}(|u_1|),..., q_{\lambda}(|u_{\Xlen}|) \big)^{\top}$.
$sgn(u) = \big( sgn(u_1),...,sgn(u_{\Xlen}) \big)^\top$ is the sign vector of $u$ with $sgn(u_j) = I(u_j>0)-I(u_j<0)$ if $u_j \neq 0$ and $sgn(u_j) = 0$ if $u_j=0$. $q_{\lambda}(|u|)sgn(u)$ is the element-wise product of $q_{\lambda}(|u|)$ and the sign function, $sgn(u)$. Throughout the paper, we use the superscript ``$p$'' to denote ``penalized''.

Different values of the tuning parameters, $\lambda_\eta$ and $\lambda_\mu$, allow for different amounts of shrinkage for parameters in the weighting models and in the outcome models. \cite{johnson2008penalized} showed that both the SCAD and the aLASSO penalty functions achieve the aforementioned three properties in the introduction: i) selection consistency, ii) producing sparse solution, and iii) preserving large coefficients. This is also called the ``oracle property'', that is, the penalized estimator is asymptotically equivalent to the oracle estimator obtained by only estimating non-zero coefficients without penalization. In this paper, we choose the SCAD penalty, although our framework is able to accommodate other penalty functions such as aLASSO. More specifically, we have 
\begin{equation*}
    q_{\lambda}(|u|) = \lambda \bigg\{ I(|u|<\lambda) + \frac{(a\lambda-|u|)_+}{(a-1)\lambda}I(|u|\geq\lambda) \bigg\},
\end{equation*}
\noindent where $\lambda>0$ is the tuning parameter. As recommended in \cite{fan2001variable}, we fix the tuning parameter $a$ as 3.7 since this choice works similarly to that chosen by generalized cross-validation method. In addition,
\begin{equation*}
   (a\lambda-|u|)_+ =
   \begin{cases}
   a\lambda-|u|, \ {\rm if}\ a\lambda-|u|\geq 0 \\
   0, \ {\rm if}\ a\lambda-|u|<0 
   \end{cases}.
\end{equation*}

Let $\pSol$ denote the solution obtained from solving the penalized estimating equation \eqref{Up_eq}. Our proposed DR estimator of the ATE is:
\begin{align}
    \hat{\theta}_{\rm DR}(\pSol) & = \frac{1}{N} \sum_{i=1}^N \Big[ 
    \Big\{ I_{A,i} d_{A,i} \EYT(X_i;\hat{\beta}^p) + \frac{I_{B,i} T_i}{\pi_B(X_i;\hat{\alpha}^p) \pi_T(X_i;\hat{\tau}^p)}\big(Y_i-\EYT(X_i;\hat{\beta}^p) \big) \Big\} \nonumber\\
    & - \Big\{ I_{A,i} d_{A,i} \EYC(X_i;\hat{\gamma}^p) + \frac{I_{B,i} (1- T_i)}{\pi_B(X_i;\hat{\alpha}^p) \big(1-\pi_T(X_i;\hat{\tau}^p)\big)} \big(Y_i-\EYC(X_i;\hat{\gamma}^p) \big) \Big\} \Big]. 
    \label{pDRestimator}
\end{align}

\section{Computation}

In this section, we introduce the numerical algorithms or computational procedures for solving the penalized estimating equation \eqref{Up_eq}. Our algorithm uses a local quadratic approximation for the SCAD penalty function \citep{fan2001variable} and uses an iterative method combined with the Newton-Raphson algorithm to obtain the solution. To be computationally efficient and to avoid inverting large matrices, we do not jointly solve for $\omega$. Instead, as guaranteed by Theorem \ref{consist_ATE}, we can fix $\mu$ when solving for $\eta$ and vice versa. Therefore, we separate the optimization for $\eta$ and $\mu$. The two separate objective functions are:

\begin{equation}\label{Up_w}
    \overline{O}^p(\eta) = 
    \begin{pmatrix}
    \partial_{\beta} \phifun(\eta,\Tilde{\mu}) \\
    \partial_{\gamma} \phifun(\eta,\Tilde{\mu})
    \end{pmatrix} + q_{\lambda_\eta}(|\eta|)sgn(\eta),
\end{equation}
and
\begin{equation}\label{Up_outcome}
    \overline{Q}^p(\mu) = 
    \begin{pmatrix}
    \partial_{\alpha} \phifun(\Tilde{\eta}, \mu) \\
    \partial_{\tau} \phifun(\Tilde{\eta}, \mu)
    \end{pmatrix} + q_{\lambda_\mu}(|\mu|)sgn(\mu),
\end{equation}

\noindent for some fixed $\Tilde{\mu}$ and $\Tilde{\eta}$. Following \cite{johnson2008penalized}, the objective function for solving \eqref{Up_w} at the $(k+1)$-th iteration is:
\begin{equation*}\label{obj_fun_eta}
    \widehat{\eta}^{(k+1)} = \argmin_{\eta} \| \overline{O}(\eta) + \Sigma_{\lambda_\eta}(\widehat{\eta}^{(k)})\eta  \|,
\end{equation*}

\noindent where $\Sigma_{\lambda_\eta} (\eta) = {\rm diag}(q_{\lambda_\eta}(|\alpha_1|)/\epsilon + |\alpha_1|,...,q_{\lambda_\eta}(|\tau_1|)/\epsilon + |\tau_1|,...)$ is a $2\Xlen \times 2\Xlen$ diagonal matrix; $\epsilon$ is a small number. We choose $\epsilon = 10^{-6}$ in our implementation.

We use the Newton-Raphson algorithm to update $\eta$ as:
\begin{equation*}
    \eta^{(k+1)} = \eta^{(k)} - \bigg\{ \triangledown(\eta^{(k)})+\Sigma_{\lambda_\eta}(\eta^{(k)}) \bigg\}^{-1} \bigg\{ \overline{O}(\eta^{(k)}) + \Sigma_{\lambda_\eta}(\eta^{(k)})\eta^{(k)} \bigg\}, 
\end{equation*}
\noindent where
\begin{align*}
    \triangledown(\eta)  & = \partial_{\eta^{\top}} \overline{O} (\eta) = \begin{pmatrix}
    \partial_{\alpha^\top} \partial_{\beta} \phifun(\eta,\Tilde{\mu}) &  
    \partial_{\tau^\top} \partial_{\beta} \phifun(\eta,\Tilde{\mu}) \\
    \partial_{\alpha^\top} \partial_{\gamma} \phifun(\eta,\Tilde{\mu}) &
    \partial_{\tau^\top} \partial_{\gamma} \phifun(\eta,\Tilde{\mu})
    \end{pmatrix}.
\end{align*}

\noindent The algorithm for solving \eqref{Up_outcome} is similar to \eqref{Up_w} except that $\triangledown(\eta)$ is replaced by $\triangledown(\mu)$ and
\begin{align*}
    \triangledown(\mu)  & = \partial_{\mu^{\top}} \overline{Q} (\mu) = \begin{pmatrix}
    \partial_{\beta^\top} \partial_{\alpha} \phifun(\Tilde{\eta}, \mu) &  
    \partial_{\gamma^\top} \partial_{\alpha} \phifun(\Tilde{\eta}, \mu) \\
    \partial_{\beta^\top} \partial_{\tau} \phifun(\Tilde{\eta}, \mu) & 
    \partial_{\gamma^\top} \partial_{\tau} \phifun(\Tilde{\eta}, \mu)
    \end{pmatrix}.
\end{align*}

\noindent For example, Under the linear additive working models, $\triangledown(\eta) $ and $\triangledown(\mu) $ have the following forms:

{\small
\begin{align*}
    \triangledown(\eta)  
    &= \frac{1}{N}\sum_{i=1}^N \begin{pmatrix}
     \frac{I_{B,i}T_i \EYT^{(1)}(X_i^{\top}\Tilde{\beta}) \pi_B^{(1)}(X_i^{\top}\alpha)}{\pi_B^2(X_i^{\top}\alpha)\pi_T(X_i^{\top}\tau) }X_i X_i^{\top} 
    & 
    \frac{I_{B,i}T_i \EYT^{(1)}(X_i^{\top}\Tilde{\beta}) \pi_T^{(1)}(X_i^{\top}\tau)}{\pi_B(X_i^{\top}\alpha)\pi_T^2(X_i^{\top}\tau) } X_i X_i^{\top}\\
    \frac{- I_{B,i}(1-T_i) \EYC^{(1)}(X_i^{\top}\Tilde{\gamma}) \pi_B^{(1)}(X_i^{\top}\alpha)}{\pi_B^2(X_i^{\top}\alpha)\big(1-\pi_T(X_i^{\top}\tau) \big) } X_i X_i^{\top}
    & 
     \frac{I_{B,i}(1-T_i) \EYC^{(1)}(X_i^{\top}\Tilde{\gamma}) \pi_T^{(1)}(X_i^{\top}\tau)}{\pi_B(X_i^{\top}\alpha) \big(1-\pi_T(X_i^{\top}\tau)\big)^2 }  X_i X_i^{\top}
    \end{pmatrix} ,
    \end{align*}
}

{\small
\begin{align*}
    \triangledown(\mu)
    &= \frac{1}{N}\sum_{i=1}^N\begin{pmatrix}
      \frac{I_{B,i}T_i \EYT^{(1)}(X_i^{\top}\beta) \pi_B^{(1)}(X_i^{\top}\Tilde{\alpha})}{\pi_B^2(X_i^{\top}\Tilde{\alpha})\pi_T(X_i^{\top}\Tilde{\tau}) } X_i X_i^{\top}
    & 
     \frac{- I_{B,i}(1-T_i)\EYC^{(1)}(X_i^{\top}\gamma) \pi_B^{(1)}(X_i^{\top}\Tilde{\alpha})}{\pi_B^2(X_i^{\top}\Tilde{\alpha}) \big(1-\pi_T(X_i^{\top}\Tilde{\tau}) \big) }  X_i X_i^{\top} \\
     \frac{I_{B,i}T_i \EYT^{(1)}(X_i^{\top}\beta) \pi_T^{(1)}(X_i^{\top}\Tilde{\tau})}{\pi_B(X_i^{\top}\Tilde{\alpha})\pi_T^2(X_i^{\top}\Tilde{\tau}) } X_i X_i^{\top}
    & 
     \frac{I_{B,i}(1-T_i) \EYC^{(1)}(X_i^{\top}\gamma) \pi_T^{(1)}(X_i^{\top}\Tilde{\tau})}{\pi_B(X_i^{\top}\Tilde{\alpha}) \big(1-\pi_T(X_i^{\top}\Tilde{\tau}) \big)^2 } X_i X_i^{\top}
    \end{pmatrix}.
\end{align*}
}

We use five-fold cross validation to choose the tuning parameters, $\lambda_\eta$ and $\lambda_\mu$. The tuning parameters are chosen to minimize the corresponding loss function:
\begin{align*}
    Loss(\lambda_\eta) &= ||\partial_{\beta} \phifun(\eta,\Tilde{\mu})||^2 + ||\partial_{\gamma} \phifun(\eta,\Tilde{\mu})||^2; \\
    Loss(\lambda_\mu) &= ||\partial_{\alpha} \phifun(\Tilde{\eta},\mu)||^2 + ||\partial_{\tau} \phifun(\Tilde{\eta},\mu)||^2.
\end{align*}
The above is summarized in Algorithm \ref{alg}, which works for most types of  outcome. For continuous outcome with a linear model, as a special case, we do not need to loop from Line 2 to Line 7 since $\EYT^{(1)}(X_i;\beta)=\EYC^{(1)}(X_i;\gamma)=1$. 

\begin{algorithm}[H] 
\caption{}\label{alg}
\begin{algorithmic}[1]
\State \textbf{Input:} $ k=0, \xi=1, \eta^{(k)}, \mu^{(k)}$;
\While{$\xi \geq 10^{-2}$}
 \State Obtain $\eta^{(k+1)}$ by optimizing $\overline{O}^p(\eta)$ at $\mu=\mu^{(k)}$ in \eqref{Up_w};
 \State Obtain $\mu^{(k+1)}$ by optimizing $\overline{Q}^p(\mu)$ at $\eta = \eta^{(k+1)}$ in \eqref{Up_outcome};
 \State $\xi = max(||\eta^{(k+1)} - \eta^{(k)}||, ||\mu^{(k+1)} - \mu^{(k)}||)$
 \State $k=k+1$;
\EndWhile
\State \textbf{Output:} $\eta^{(k+1)}, \mu^{(k+1)}$
\end{algorithmic}
\end{algorithm}

\section{Asymptotic properties}

\subsection{Asymptotic properties of the penalized estimating equation}

We establish the asymptotic properties of our proposed penalized estimating equation and the DR estimation method. Let $n=n_A + n_B$ be the sample size combining two datasets. Following \cite{chen2020doubly}, we make the assumption about the relationship between the total population $N$ and the sample sizes $n_A$ and $n_B$:

\begin{assumption} \label{assumption_N} 
The population size $N$ and the sample sizes $n_A$ and $n_B$ satisfy $\lim_{N\rightarrow\infty} n_A/N = f_A \in (0,1)$ and $\lim_{N\rightarrow\infty} n_B/N = f_B \in (0,1)$.
\end{assumption}
Assumption \ref{assumption_N} implies that $n_A$ and $n_B$ increase as fast as $N$, and therefore, we do not need to distinguish among $O_p(n_A^{-1/2}),O_p(n_B^{-1/2}),O_p(n^{-1/2})$ and $O_p(N^{-1/2})$. Define the set $\mathcal{M}_\omega = \big\{j:\omega^{\ast}_j \neq 0,\ j\in \{1,2,...,4\Xlen \} \big\}$ and $\lambda_\omega = max (\lambda_\eta, \lambda_\mu)$. We assume that the following regularity conditions hold, which are commonly seen in the penalized estimating equation literature \citep{johnson2008penalized}.

\begin{assumption} \label{A_reg} Regularity conditions.
\begin{enumerate}[label={(\alph*)}, ref={(\alph*)}] 
\item \label{A_reg_a}%
    There exists a non-singular matrix $\Omega$ such that for any given constant $M$, 
\begin{equation*}
    \sup_{|\omega-\omega^\ast|\leq Mn^{-1/2}} | \sqrt{n} \Ufun(\omega) - \sqrt{n} \Ufun(\omega^\ast) - \sqrt{n} \Omega (\omega - \omega^\ast) | = o_p(1),
\end{equation*}
\noindent \hspace{0.5cm} $\sqrt{n} \Ufun(\omega^\ast) \overset{\text{d}}{\sim} N(0,W)$ and $\Omega_{\mathcal{M}_\omega, \mathcal{M}_\omega}$ is non-singular;
\item \label{A_reg_b}%
    For nonzero fixed $\omega$, $\sqrt{n} q_{\lambda_\omega} (|\omega|) \rightarrow 0$;
\item\label{A_reg_c}%
    For any $M>0$, $\inf_{|\eta|<M n^{-1/2}} \sqrt{n} q_{\lambda_{\eta}} (|\eta|) \rightarrow \infty$ and $\inf_{|\mu|<M n^{-1/2}} \sqrt{n} q_{\lambda_{\mu}} (|\mu|) \rightarrow \infty$.
\end{enumerate}
\end{assumption}

Assumption \ref{A_reg}\ref{A_reg_c} implies that for any $M>0$, $\inf_{|\omega|<M n^{-1/2}} \sqrt{n} q_{\lambda_{\omega}} (|\omega|) \rightarrow \infty$. Assumption \ref{A_reg}\ref{A_reg_a} is commonly seen and used for Z-estimators \citep{wellner2013weak}. Assumptions \ref{A_reg}\ref{A_reg_b} and \ref{A_reg}\ref{A_reg_c} put restrictions on the choice of the penalty function and the regularization parameters. In particular, Assumption \ref{A_reg}\ref{A_reg_b} prevents overly shrinking $\omega$ if $\omega\neq 0$, and Assumption \ref{A_reg}\ref{A_reg_c} allows one to shrink $\omega$ to 0 if $\omega =0$ for a sparse solution. For the SCAD penalty with $a>2$, Assumption \ref{A_reg}\ref{A_reg_b} holds if $\lambda_\omega \rightarrow 0$ and Assumption \ref{A_reg}\ref{A_reg_c} holds if 
$\sqrt{n}\lambda_{\eta} \rightarrow \infty$ and $\sqrt{n}\lambda_{\mu} \rightarrow \infty$.

We now consider the solution to the penalized estimating equation, $\Ufun ^p(\omega)=0$. Following \cite{johnson2008penalized}, we provide a formal definition of the solution to the penalized estimating equation. An estimator $\Tilde{\omega}$ is called an approximate zero-crossing of $\Ufun ^p(\omega)$ if for $j=1,2,...,4\Xlen$ and $\epsilon>0$,
\begin{equation*}
\lim_{N\rightarrow\infty} P\big\{ \uplim_{\epsilon \rightarrow 0+}  n \Ufun _j^p(\Tilde{\omega} + \epsilon e_j) \Ufun _j^p(\Tilde{\omega} - \epsilon e_j) \leq 0 \big\} = 1,
\end{equation*}
\noindent where $e_j$ is the j-th canonical unit vector and $\Ufun _j^p(\cdot)$ is the j-th element in $\Ufun ^p(\cdot)$. 
Briefly speaking, the approximate zero-crossing $\Tilde{\omega}$ is the solution to the penalized estimation equation in the sense that a small change in any element in the $\Tilde{\omega}$ will change the sign of the objective function. This feature ensures that $\Tilde{\omega}$ is close enough to the exact solution of solving $\Ufun ^p(\omega)=0$.
This definition will be used in the next theorem for establishing the asymptotic properties of the penalized estimating equation.

\begin{theorem} \label{theorem_omega}

Under Assumptions \ref{assumption_N} and \ref{A_reg}, there is an approximate zero-crossing $\Tilde{\omega}$, which satisfies:
\begin{enumerate}[label={(\alph*)}, ref={(\alph*)}] 
\item \label{omega_solution} $\Ufun _j^p(\Tilde{\omega}) = o_p(1)\ {\rm and}\ \Ufun _j(\Tilde{\omega}) = o_p(1),\ for\ all\ j=1,...,4\Xlen.$
\item \label{omega_sparse} sparsity: $P(\Tilde{\omega}_{\mathcal{M}_{\omega}^c} = 0) \rightarrow 1.$
\item \label{omega_rate} root-n-consistency: $\Tilde{\omega} - \omega^{\ast} = O_p(n^{-1/2}).$
\end{enumerate}
\end{theorem}
We present the proof for Theorem \ref{theorem_omega} in Section S1.2 of the supplementary materials. Theorem \ref{theorem_omega}\ref{omega_sparse} implies that the penalized estimating equations produce sparse solutions. Theorem \ref{theorem_omega}\ref{omega_rate} shows that the estimator is root-n consistent. Overall, Theorem \ref{theorem_omega} shows that our penalized estimating equation procedure achieves the oracle property for variable selection and estimation.

\subsection{Asymptotic properties of the DR estimator}
We now show the asymptotic properties of $\hat{\theta}_{\rm DR}(\pSol)$ in \eqref{pDRestimator}. By Taylor series expansion, 
\begin{align*}
    \sqrt{n} \big\{\hat{\theta}_{\rm DR}(\pSol) - \theta_0 \big\} =&  
    \sqrt{n} \big\{\hat{\theta}_{\rm DR}(\omega^{\ast}) - \theta_0 \big\} + \\ 
    & \sqrt{n} \frac{\partial \hat{\theta}_{\rm DR} (\pSol)}{\partial \omega^{\top}} 
    (\pSol - \omega^\ast)+ 
    O_p\bigg(\sqrt{n} \norm{ \pSol - \omega^\ast}^2 \bigg),
\end{align*}
\noindent where $\hat{\theta}_{\rm DR}(\omega^\ast)$ denote the DR estimator by plugging in $\omega^\ast$. The third term on the right-hand side is $o_p(1)$ following Theorem \ref{theorem_omega}\ref{omega_rate}. The second term is the estimating function $\Ufun(\pSol)$, which is $o_p(1)$ following Theorem \ref{theorem_omega}\ref{omega_solution}. Therefore, 
\begin{equation*}
    \sqrt{n} \big\{\hat{\theta}_{\rm DR}(\pSol) - \theta_0 \big\}= \sqrt{n} \big\{ \hat{\theta}_{\rm DR}(\omega^\ast) - \theta_0 \big\} + o_p(1).
\end{equation*}
That is, $\hat{\theta}_{\rm DR}(\pSol)$ is asymptotically equivalent to $\hat{\theta}_{\rm DR}(\omega^{\ast})$. It follows that the randomness of $\pSol$ can be ignored when making the inference about $\theta_0$. 

\begin{remark} \label{remark3}
\normalfont
    The two-step approach \citep{yang2020doubly} requires that the first term of the Taylor series expansion with respect to all the nuisance parameters equals zero, i.e., $\sqrt{n} \frac{\partial \hat{\theta}_{\rm DR} (\pSol)}{\partial \omega^{\top}} 
    (\pSol - \omega^\ast)=0$. For a non-zero coefficient, this zero condition requires that the partial derivative of the DR estimator with respect this nuisance parameter in $\frac{\partial \hat{\theta}_{\rm DR} (\pSol)}{\partial \omega^{\top}}$ is zero. However, their proposed estimating equations in the second estimation step only ensure this zero condition of the partial derivative for parameters associated with selected variables. For unselected variables, the Taylor series expansion term for their parameters is not necessarily zero, for example, in case of false negatives from the variable selection procedure. 
\end{remark}

Recall that the random variables in $\hat{\theta}_{\rm DR}(\omega^{\ast})$ are $I_A, I_B, T, X$ and $Y$. By iterative expectation, we have that $\hat{\theta}_{\rm DR}(\omega^{\ast})$ is unbiased for $\theta_0$ if either $\pi_B(X;\alpha)$ and $\pi_T(X;\tau)$ or $\EYT(X;\beta)$ and $\EYC(X;\gamma)$ are correctly specified. Note that
{
\begin{align}
    E\big\{ & \hat{\theta}_{\rm DR}(\omega^{\ast}) - \theta_0\big\} = E_{X}\Big[\EYT(X_i; \beta^\ast) - \EYC(X_i;\gamma^\ast) - \theta_0 + \nonumber\\
    & E_{I_B,T|X}\Big\{ \frac{I_{B,i}T_i}{\pi_B(X_i;\alpha^{\ast}) \pi_T(X_i; \tau^{\ast})} \Big\} E_{Y(1)|X}\big\{ Y_i(1) - \EYT(X_i;\beta^{\ast}) \big\}- \nonumber \\
    & E_{I_B,T|X}\Big\{ \frac{I_{B,i} (1-T_i )}{\pi_B(X_i;\alpha^{\ast}) \big(1-\pi_T(X_i; \tau^{\ast}) \big)} \Big\} E_{Y(0)|X}\big\{ Y_i(0) - \EYC(X_i;\gamma^{\ast}) \big\} \Big]. \label{asy_bias}
\end{align}
}
If weighting models $\pi_B(X;\alpha)$ and $\pi_T(X;\tau)$ are correctly specified, equation \eqref{asy_bias} reduces to 
$E_X \big[ E_{Y(1)|X}\big\{ Y_i(1)\big\} - E_{Y(0)|X}\big\{Y_i(0)\big\} - \theta_0 \big]= E\{Y(1)-Y(0)\} - \theta_0 = 0$; If outcome models $\EYT(X;\beta)$ and $\EYC(X;\gamma)$ are correctly specified, equation \eqref{asy_bias} reduces to $E_X \big\{ \EYT(X_i;\beta^\ast) - \EYC(X_i;\gamma^\ast) - \theta_0 \big\}= E\{Y(1)-Y(0)\} - \theta_0 = 0$. Therefore, $\hat{\theta}_{\rm DR}(\omega^\ast)$ is unbiased under the double robustness condition.


To establish the asymptotic variance of $\sqrt{n} \big\{ \hat{\theta}_{\rm DR}(\omega^\ast) - \theta_0\big\}$, we decompose the total variance into two parts \citep{yang2020doubly,shao1999variance}, where the first part involves the conditional variance in sample $A$ and the second part involves the variance of the conditional expectation in sample $B$. More specifically, 
\begin{align*}
    V \Big\{\sqrt{n} \big( \hat{\theta}_{\rm DR}(\omega^\ast) - \theta_0\big) \Big\} =& n E_{I_B,T,X,Y}\Big[V_{I_A|I_B,T,X,Y}\big\{ \hat{\theta}_{\rm DR}(\omega^\ast) - \theta_0 \big\} \Big] + \\
    & n V_{I_B,T,X,Y}\Big[E_{I_A|I_B,T,X,Y} \big\{ \hat{\theta}_{\rm DR}(\omega^\ast) - \theta_0 \big\} \Big].
\end{align*}
Denote the first and the second term on the right hand side as $V_1$ and $V_2$, respectively. We have:
\begin{align}
        V_1 &= E_{X} \bigg\{ \frac{n}{N^2} \sum_{i=1}^{N}   (d_{A,i} - 1)\big( \EYT(X_i; \beta^\ast) - \EYC(X_i; \gamma^\ast) \big)^2 \bigg\} \label{V_1}
\end{align}
is the sampling variance of the Horvitz-Thompson estimator for the probability sample $A$ and 
\begin{align}\label{V_2}
    V_2 = \frac{n}{N} E_{I_B, T, X, Y} & 
    \bigg[ \frac{I_{B,i}T_i}{\pi_B(X_i;\alpha^{\ast}) \pi_T(X_i; \tau^{\ast})} \big\{ Y_i(1) - \EYT(X_i;\beta^{\ast}) \big\} -  \nonumber \\
    & \frac{I_{B,i} (1-T_i )}{\pi_B(X_i;\alpha^{\ast}) (1-\pi_T(X_i; \tau^{\ast}) )} \big\{ Y_i(0) - \EYC(X_i;\gamma^{\ast}) \big\}  + \\
    & \big\{ \EYT(X_i; \beta^\ast) - \EYC(X_i;\gamma^\ast) - \theta_0 \big\}  \bigg]^2 .
\end{align} 
\noindent The derivation of $V_1$ and $V_2$ is provided in Section S2.1 of the supplementary materials. We formally provide the asymptotic properties of $\hat{\theta}_{\rm DR}(\pSol)$ in the next theorem.

\begin{theorem}
If $\pi_B(X;\alpha)$ and $\pi_T(X;\tau)$ or $\EYT(X;\beta)$ and $\EYC(X;\gamma)$ are correctly specified,
\begin{equation*}
    \sqrt{n} \big\{\hat{\theta}_{\rm DR}(\pSol) - \theta_0 \big\} \xrightarrow{d} N(0,V),
\end{equation*}
\noindent where $V = \lim(V_1 +V_2)$ and $V_1, V_2$ are defined in \eqref{V_1} and \eqref{V_2}.
\end{theorem}

$V_1$ can be estimated using the design-consistent estimator in survey sampling literature \citep{kott1990estimating,deville1992calibration,deville1994variance,breidt2000local}:
\begin{equation*}
    \hat{V}_1 = \frac{n}{N^2} \sum_{i=1}^N I_{A,i}d_{A,i}(d_{A,i}-1) \big(\EYT(X_i; \hat{\beta}^p) - \EYC(X_i;\hat{\gamma}^p)\big)^2.
\end{equation*}
Following the law of large numbers, $V_2$ can be consistently estimated by the sample mean in sample $B$. 
A more detailed discussion about $\hat{V}_1$ being design-consistent and the expression of $\hat{V}_2$ is provided in Section S2 of the supplementary materials.

\section{Simulation}
\subsection{Simulation setup}
In this section, we evaluate the performance of our proposed penalized estimating equation procedure and the DR estimator $\hat{\theta}_{\rm DR}(\pSol)$ in \eqref{pDRestimator}. We first generate the complete data of the target population $\{(X_i, T_i, Y_i), i=1,...,N\}$ with $N=50,000$. Specifically, we generate $X_i = (1,X_{i,1},X_{i,2},...,X_{i,50})$ containing the intercept and 50 covariates. Covariates, $X_{i,1},X_{i,2},...,X_{i,50}$, are generated independently from a standard normal distribution. We then generate the treatment indicator from a linear or non-linear treatment model ($TM$):
\begin{enumerate}[label=(\alph*)]
    \item ${\rm logit}\big(P(T_i=1|X_i) \big) = -1-0.5X_{i,1}-0.5X_{i,2}-0.5X_{i,3}$;
    \item ${\rm logit}\big(P(T_i=1|X_i) \big) = -1-0.5X_{i,1}^2-0.5X_{i,2}^2-0.5I(X_{i,3}>0.5)$.
\end{enumerate}
\noindent The continuous outcomes are generated from a linear or non-linear outcome model ($OM$):
\begin{enumerate}[label=(\alph*)]
    \item $Y_i = 1 + T_i + X_{i,1} + 2T_i\times X_{i,1} + X_{i,2} + X_{i,3} + X_{i,4} + X_{i,5} + \epsilon_i$, where $\epsilon_i \sim N(0,1)$;
    \item $Y_i = 1 + T_i + |X_{i,1}| + 2T_i\times |X_{i,1}| + |X_{i,2}| + |X_{i,3}| + |X_{i,4}| + |X_{i,5}| + \epsilon_i$, where $\epsilon_i \sim N(0,1)$.
\end{enumerate}

After simulating the complete data for the target population, we draw a probability sample $A$, where the probability for each individual in the target population to be selected into sample $A$ is 0.02. The expected sample size for sample $A$, $n_A$, is 1,000. In the remaining population, we select a non-probability sample $B$ according to the following selection model ($SM$):
\begin{enumerate}[label=(\alph*)]
    \item ${\rm logit}\big(P(I_{B,i}=1|X_i) \big) = -2.3+0.5X_{i,1}+0.5X_{i,2}+0.5X_{i,3}$;
    \item ${\rm logit}\big( P(I_{B,i}=1|X_i) \big) = -3.2+\big(I(X_{i,1}>1)+I(X_{i,2}>1)+I(X_{i,2}>1)\big)^2$.
\end{enumerate}

The sample size for sample $B$, $n_B$ is about 5,500. We have eight different combinations for data generation, where the treatment indicator, $T$, outcome, $Y$, and the selection indicator, $I_B$, are generated from a linear or non-linear model (Cases 1-8 listed in \textbf{Table \ref{selection_result_cont}}). When applying our proposed method, we always specify a linear form of covariates so that models are misspecified if the underlying true model is non-linear. We repeat each simulation scenario 500 times and the results are averaged over 500 replications.

We compare the performance of our proposed DR estimator, $\hat{\theta}_{\rm DR}(\pSol)$, to other estimators listed below. We do not compare our method to a two-step estimator implied from \cite{yang2020doubly} since their work focuses on estimating the population mean instead of the ATE. 
\begin{enumerate}[label=(\alph*)]
    \item $\hat{\theta}_{\rm NAIVE}$, the naive estimator using only the non-probability sample $B$. That is,
    \begin{equation*}
        \hat{\theta}_{\rm NAIVE} = \big( \sum_{i=1}^N I_{B,i} \big)^{-1} 
        \Big\{ \sum_{i=1}^N I_{B,i} \big(\EYT(X_i^{\top}\hat{\beta}) - \EYC(X_i^{\top}\hat{\gamma})\big) \Big\},
    \end{equation*}
    where $\hat{\beta}, \hat{\gamma}$ are obtained by fitting a regression model on the non-probability sample;
    \item $\hat{\theta}_{\rm ORACLE}$, the DR estimator in \eqref{DR_estimator_existing}, where $\hat{\alpha},\hat{\tau},\hat{\beta},\hat{\gamma}$ are obtained based on the set of true predictors; 
    \item $\hat{\theta}_{\rm OR}$, the OR estimator in \eqref{OR_estimator},
    where $\hat{\beta},\hat{\gamma}$ are obtained from likelihood-based estimating equations with the SCAD penalty; 
    \item $\hat{\theta}_{\rm IPW}$, the IPW estimator in \eqref{IPW_estimator},
    where $\hat{\alpha},\hat{\tau}$ are obtained from the calibration-based estimating equation with the SCAD penalty. 
\end{enumerate}

We also perform simulation studies for binary outcomes, following the above data generation procedure with only modification to the outcome model. Models for generating binary outcomes are described in Section S4 in the supplementary materials. 

\subsection{Simulation results}

We present the simulation results in terms of four aspects: (a) sensitivity and specificity of variable selection from our proposed penalized estimating equation procedure. Sensitivity and specificity quantify how well the variable selection procedure identifies the underlying important variables and removes the irrelevant variables. (b) Mean squared error (MSE) for the non-null and null coefficients from the penalized estimating equation. (c) Estimates of ATE using our proposed DR estimator, $\hat{\theta}_{\rm DR}(\pSol)$, and other competing estimators. (d) The coverage probability of the 95\% confidence interval (CI). Metrics for the penalization procedure are defined as:
\begin{equation*}
    {\rm sensitivity} = R^{-1} \sum_{r=1}^R \bigg\{ \frac{\sum_{i=1}^d I(\beta_j\neq0)I(\hat{\beta}_j^{(r)}\neq0)}{\sum_{i=1}^d I(\beta_j\neq0)} \bigg\},
\end{equation*}
\begin{equation*}
    {\rm specifity} = R^{-1} \sum_{r=1}^R \bigg\{ \frac{\sum_{i=1}^d I(\beta_j=0)I(\hat{\beta}_j^{(r)}=0)}{\sum_{i=1}^d I(\beta_j=0)} \bigg\},
\end{equation*}
\begin{equation*}
    {\rm MSE\ for\ non-null\ coefficients} = R^{-1} \sum_{r=1}^{R}\bigg\{ \frac{\sum_{j=1}^{d} I(\beta_j\neq0)(\hat{\beta}_j^{(r)}-\beta_j)^2}{\sum_{j=1}^{d} I(\beta_j\neq0)}\bigg\},
\end{equation*}
\begin{equation*}
    {\rm MSE\ for\ null\ coefficients} = R^{-1} \sum_{r=1}^{R}\bigg\{ \frac{\sum_{j=1}^{d} I(\beta_j=0)(\hat{\beta}_j^{(r)}-\beta_j)^2}{\sum_{j=1}^{d} I(\beta_j=0)}\bigg\},
\end{equation*}
\noindent where $r$ is the indicator of the r-th simulation replication.

In \textbf{Table \ref{selection_result_cont}}, we present the selection results in terms of sensitivity and specificity for continuous outcomes using our proposed penalized estimating equation procedure. The sensitivity and specificity are close to 1 when models are correctly specified. As expected, the sensitivity, in general, reduces when models are misspecified. For example, when the treatment model is misspecified in Case 3, the proposed method on average only selects 57.1\% of the true predictors in the treatment model. When the outcome models are misspecified, similar changes of sensitivity and specificity for $\beta$ and $\gamma$ are observed. For example, when the outcome model is misspecified in Case 5, only 22.5\% and 23.9\% of true signals in $\beta$ and $\gamma$ are selected, respectively. 

\textbf{Table \ref{MSE_results_cont}} shows the MSE for the non-null and null coefficients with continuous outcomes when all models are correctly specified. The MSE for both non-null and null coefficients in all models are small, indicating that our penalized estimating equation procedure yields nearly unbiased estimates for the nuisance parameters.

{
\begin{table}[tb]
\centering
\caption{Sensitivity (sens) and specificity (spec) of our proposed penalized estimating equation for continuous outcomes based on 500 simulation replications. The subscript $c$ stands for the correct specification of the outcome model ($OM$), selection model ($SM$), and treatment model ($TM$) while the subscript $m$ stands for model misspecification.}
\begin{tabular}{p{3.8cm}rrrrrrrr}
  \hline
  & \begin{tabular}[c]{@{}c@{}} $\alpha$ \\ sens \end{tabular} & \begin{tabular}[c]{@{}c@{}} $\alpha$ \\ spec \end{tabular} & \begin{tabular}[c]{@{}c@{}} $\tau$ \\ sens \end{tabular} & \begin{tabular}[c]{@{}c@{}} $\tau$ \\ spec \end{tabular} & \begin{tabular}[c]{@{}c@{}} $\beta$ \\ sens \end{tabular} & \begin{tabular}[c]{@{}c@{}} $\beta$ \\ spec \end{tabular} & \begin{tabular}[c]{@{}c@{}} $\gamma$ \\ sens \end{tabular} & \begin{tabular}[c]{@{}c@{}} $\gamma$ \\ spec \end{tabular} \\ 
Case 1: $OM_c, SM_c, TM_c$ & 1 & 0.995 & 1 & 0.982 & 1 & 0.992 & 1 & 0.998 \\ 
  Case 2: $OM_c, SM_m, TM_c$ & 1 & 1 & 1 & 0.997 & 1 & 0.956 & 1 & 0.966 \\ 
  Case 3: $OM_c, SM_c, TM_m$ & 1 & 0.989 & 0.571 & 0.992 & 1 & 0.978 & 1 & 0.999 \\ 
  Case 4: $OM_c, SM_m, TM_m$ & 1 & 0.999 & 0.853 & 1 & 1 & 0.98 & 1 & 0.999 \\ 
  Case 5: $OM_m, SM_c, TM_c$ & 1 & 0.999 & 1 & 0.997 & 0.225 & 0.992 & 0.239 & 0.997 \\ 
  Case 6: $OM_m, SM_m, TM_c$ & 1 & 1 & 1 & 0.997 & 0.248 & 0.991 & 0.269 & 0.996 \\ 
  Case 7: $OM_m, SM_c, TM_m$ & 1 & 0.989 & 0.571 & 0.992 & 0.199 & 0.982 & 0.184 & 1 \\ 
  Case 8: $OM_m, SM_m, TM_m$ & 1 & 0.989 & 0.692 & 0.993 & 0.24 & 0.984 & 0.197 & 1 \\ 
  \hline
\end{tabular}
\label{selection_result_cont}
\end{table}
}

\begin{table}[tb]
\centering
\caption{Mean squared error (MSE) for non-null and null coefficients with continuous outcomes using our proposed penalized estimating equation based on 500 simulation replications. MSE is evaluated when all models are correctly specified.}
\label{MSE_results_cont}
\begin{tabular}{ M{1.2cm}M{1.2cm}M{1.2cm}M{1.2cm}M{1.2cm}M{1.2cm}M{1.2cm}M{1.2cm}}
  \hline
  \begin{tabular}[c]{@{}c@{}} $\alpha$ \\ $non-null$ \end{tabular} & \begin{tabular}[c]{@{}c@{}} $\alpha$ \\ $Null$ \end{tabular} & \begin{tabular}[c]{@{}c@{}} $\tau$ \\ $non-null$ \end{tabular} & \begin{tabular}[c]{@{}c@{}} $\tau$ \\ $Null$ \end{tabular} & \begin{tabular}[c]{@{}c@{}} $\beta$ \\ $non-null$ \end{tabular} & \begin{tabular}[c]{@{}c@{}} $\beta$ \\ $Null$ \end{tabular} & \begin{tabular}[c]{@{}c@{}} $\gamma$ \\ $non-null$ \end{tabular} & \begin{tabular}[c]{@{}c@{}} $\gamma$ \\ $Null$ \end{tabular} \\ 
1.13e-01 & 1.19e-04 & 1.66e-02 & 1.59e-03 & 1.03e-02 & 1.22e-04 & 1.86e-02 & 1.19e-04 \\  \hline
\end{tabular}
\end{table}

\begin{table}[tb]
\centering
\caption{Coverage properties of the 95\% confidence interval for continuous outcomes based on 500 replications: empirical coverage rate and empirical coverage rate $\pm$ $2\times$Monte Carlo standard error. The subscript $c$ stands for the correct specification of the outcome model ($OM$), selection model ($SM$), and treatment model ($TM$) while the subscript $m$ stands for model misspecification. }
\begin{tabular}{lr}
  \hline
Case 1: $OM_c, SM_c, TM_c$ & 0.962 (0.945,0.979) \\ 
  Case 2: $OM_c, SM_m, TM_c$ & 0.946 (0.926,0.966) \\ 
  Case 3: $OM_c, SM_c, TM_m$ & 0.964 (0.947,0.981) \\ 
  Case 4: $OM_c, SM_m, TM_m$ & 0.964 (0.947,0.981) \\ 
  Case 5: $OM_m, SM_c, TM_c$ & 0.966 (0.950,0.982) \\ 
  Case 6: $OM_m, SM_m, TM_c$ & 0.058 (0.037,0.079) \\ 
  Case 7: $OM_m, SM_c, TM_m$ & 0 (0,0) \\ 
  Case 8: $OM_m, SM_m, TM_m$ & 0 (0,0) \\ 
    \hline
\end{tabular}
\label{coverage_results_continuous}
\end{table}

\begin{figure}[tb]
    \centering
    \includegraphics[width = 12cm]{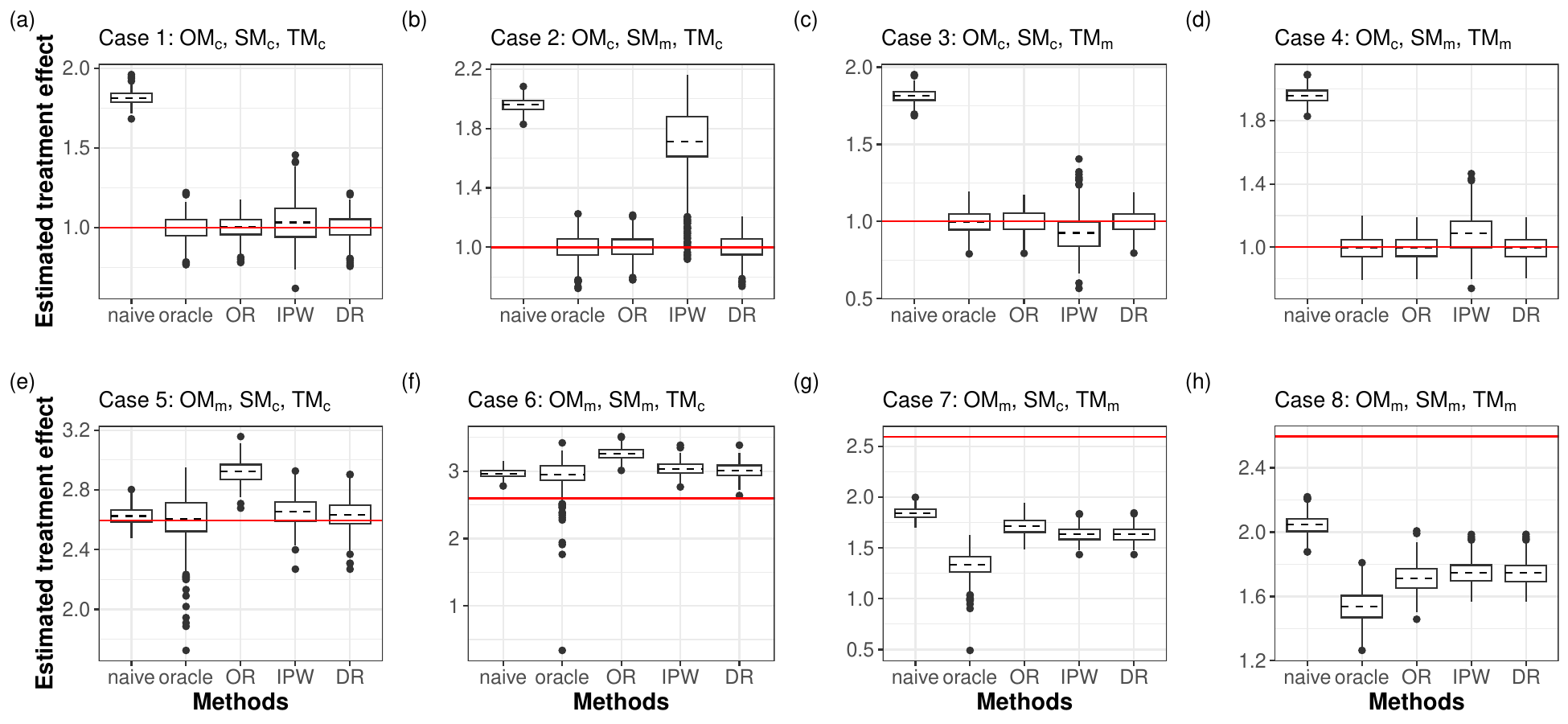}
    \caption{Estimated average treatment effect using our proposed DR estimator and four other competing estimators with continuous outcomes based on 500 replications. The subscript $c$ stands for the correct specification of the outcome model ($OM$), selection model ($SM$), and treatment model ($TM$) while the subscript $m$ stands for model misspecification. The red horizontal line indicates the true average treatment effect.}
    \label{estimation_ATE_figure_cont}
\end{figure}

In \textbf{Figure \ref{estimation_ATE_figure_cont}}, we present the ATE estimates for continuous outcomes using our proposed DR estimator and four other competing estimators. The naive estimator $\hat{\theta}_{\rm NAIVE}$, which ignores the sampling bias in the non-probability sample, is greatly biased across all simulation scenarios. The oracle estimator, $\hat{\theta}_{\rm ORACLE}$, which uses true predictors, is consistent if either the weighting models or the outcome models are correctly specified. On the other hand, the OR estimator, $\hat{\theta}_{\rm OR}$, is unbiased only when the outcome models are correctly specified; and the IPW estimator, $\hat{\theta}_{\rm IPW}$ is nearly unbiased when the weighting models are correctly specified. Our proposed DR estimator, $\hat{\theta}_{\rm DR}(\pSol)$, performs similarly to the oracle estimator and is unbiased under the double robustness condition, i.e., if the outcome models or the weighting models are correctly specified. 

\textbf{Table \ref{coverage_results_continuous}} shows the coverage properties of the 95\% CI. Under the double robustness condition, the coverage rates are close to the nominal level. For instance, the coverage rate of our proposed DR estimator under Case 1 where all models are correctly specified is 0.962, compared to 0.946, 0.964, 0.964 in Cases 2-4, respectively, where only outcome models are correctly specified, and compared to 0.966 in Case 5 where only models for selection and treatment are correctly specified. Results for the binary outcomes are presented in \textbf{Tables S1-S3} and \textbf{Figure S1} in the supplementary materials. The main observations for binary outcomes are similar to those for continuous outcomes. Thus, we do not repeat it here.

As we previously mentioned in Section 2, an alternative DR estimator of the ATE as shown in \eqref{DRestimator2} is to directly jointly model $P(I_B=1,T=t|X)$. However, if we directly apply the alternative DR estimator of the ATE under the above data generation mechanism in Cases 1-8, the working models $w_1(X,\delta_1)$ for $P(I_B=1,T=1|X)$ and $w_0(X,\delta_0)$ for $P(I_B=1,T=0|X)$ are likely always misspecified (\textbf{Figures S2-S3}). Therefore, we conducted additional simulation studies using the alternative DR estimator for the purpose of completeness. Details about data generation can be found in the supplementary materials Section S5.1. In general, the primary observations of the joint approach are consistent with those of the conditional approach discussed previously (\textbf{Tables S4-S9}, \textbf{Figures S4-S7}). In short, the penalized estimating equation procedure has high sensitivity and specificity for variable selection and low MSE for coefficient estimation if the corresponding working models are correctly specified. The alternative DR estimator of the ATE based on the joint models of $P(I_B=1,T=t|X)$ is consistent under the doubly robust condition, i.e., if the outcome models $\EYT(X,\beta), \EYC(X,\gamma)$ or the weighting models $w_1(X,\delta_1), w_0(X,\delta_0)$ are correctly specified, but not necessarily both. The estimated CI also achieves 95\% coverage rates.

\section{Analysis of the Michigan Genomics Initiative (MGI) data}

We analyze two datasets from the MGI and the 2017-2018 NHANES. MGI is an EHR-linked bio-respiratory within Michigan Medicine of over 80,000 patients. The EHR dataset from MGI is a non-probability sample containing patient diagnoses, demographics, lifestyle and behavioral risk factors, and laboratory results. The 2017-2018 NHANES dataset is a probability sample representing the US adult population. It contains 5,569 adult participants (age $\geq$ 20) with known survey weights. 

\begin{figure}[htbp]
    \centering
    \includegraphics[width = 10cm]{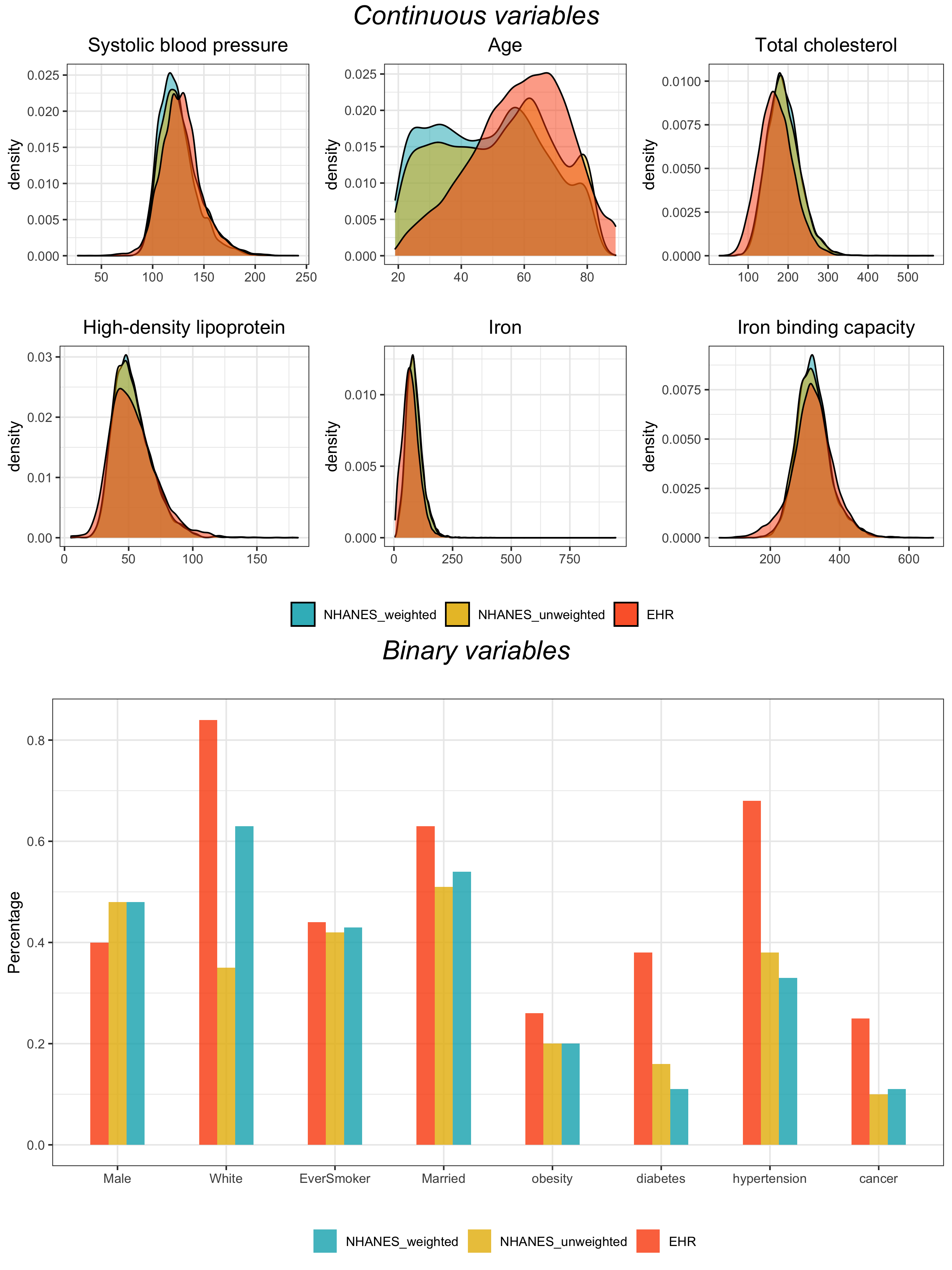}
    \caption{Covariate distribution in the EHR data from MGI, in the 2017-2018 NHANES, and in the US population (weighted distribution using NHANES survey weights). }
    \label{density_plot}
\end{figure}

The treatment variable of interest is severe obesity, defined as BMI $\geq35$, with 1 indicating severe obesity. We focus on two outcome variables: systolic blood pressure (continuous outcome) and hypertension (binary outcome). Eleven covariates are observed in both datasets, including age, gender, race (categorized as white or non-white), smoking history (smoked before or never smoked), marital status, diabetes, cancer, total cholesterol levels, high-density lipoprotein levels, as well as measurements of iron and iron binding capacity. To highlight the heterogeneity of the two datasets, \textbf{Figure \ref{density_plot}} shows the distribution of covariates from the EHR data and the NHANES data, and the US population distribution constructed by the NHANES survey weights. As expected in a hospital-based dataset, EHR patients in the MGI tend to be older and have a greater burden of disease compared to NHANES participants and the US population. After removing all missing values, we have 13,112 patients in the EHR data and 4,569 participants in the NHANES data with complete observations. 

Before the analysis, we log-transform right-skewed variables, including total cholesterol, high-density lipoprotein, and iron. We standardize all continuous variables by subtracting the mean and dividing by the standard deviation of each variable after combining two datasets. Our models also include 45 pairwise interaction terms among the set of covariates. When constructing the interaction terms, we omit interactions related to race, since 84\% of the EHR patients are white. In total, we have 56 covariates and interaction terms. 

We apply our proposed DR estimator in \eqref{pDRestimator} with the penalized estimating equation procedure in \eqref{Up_eq} to estimate the ATE of severe obesity on the outcome of interest. In addition, we calculate various other ATE estimators, as outlined in \textbf{Table S10} using different data sources. This analysis allows us to examine three main aspects: (a) the impact of ignoring both confounding bias and selection bias, where we exclusively use the EHR data; (b) the impact of ignoring selection bias only, where we exclusively use the EHR data; and (c) the advantage of penalization, where we use both the EHR data and the NHANES data. 
Furthermore, we compute the ATE estimators exclusively based on the NHANES data. We note that in this particular data example, this is feasible since both the treatment and the outcome are observed in the NHANES. The DR estimator using the NHANES data exclusively can be considered as the``oracle'' estimator for comparison. The variance estimation follows from the theory of M-estimation \citep{stefanski2002calculus} and more details about variance derivation are provided in Section S3 of the supplementary materials.

\textbf{Figure \ref{SBP_ATE_plot}} shows the ATE estimates with the 95\% CI from our proposed method and other competing methods for the continuous outcome. Out of a total of 56 covariates, our penalized estimating equation procedure selects 2 variables in the selection model, 11 in the treatment model, 12 in the outcome model for the treatment arm, and 11 in the outcome for the control arm. The oracle ATE estimate is 6.32 (4.22, 8.41) and our proposed estimate is 5.20 (3.87, 6.54). Compared to OR and DR estimators, the IPW-type estimators have wider CIs. Estimators that solely use the EHR data, i.e., the unadjusted mean-difference estimator and the OR/IPW/DR estimators, are biased due to ignoring selection bias. For example, the ATE estimate obtained from the DR estimator based exclusively on the EHR data is 4.18 (3.32, 5.04), compared to the oracle estimate 6.32 (4.22, 8.41).
Integrating the NHNAES data with the EHR data effectively addresses the selection bias issue. The ATE estimate from the DR estimator using combined data is 5.20 (3.87,6.54) with penalization and 5.81 (4.28, 7.34) without penalization, respectively. 
The DR estimator using both datasets has a smaller standard error (0.68 with penalization and 0.78 without penalization) than the oracle estimator (1.07) due to an increase of sample size. 
In addition, our proposed DR estimator of the ATE with the penalized estimating equation procedure has a smaller standard error of 0.68 than that without the penalization procedure of 0.78 due to removing irrelevant variables.

Results of the ATE estimates for the binary outcome are shown in \textbf{Figure S8}. Using our proposed method, we find that the ATE of severe obesity on hypertension is 0.17 (0.13, 0.21). In addition, we provide ATE estimates via directly estimating the joint probability of $P(I_B=1,T=t|X)$ (\textbf{Table S11}). The ATE estimate from the alternative DR estimator using the joint approach are almost identical to that from the conditional approach for both the continuous and binary outcomes.

\begin{figure}[htbp]
    \centering
    \includegraphics[width = 12cm]{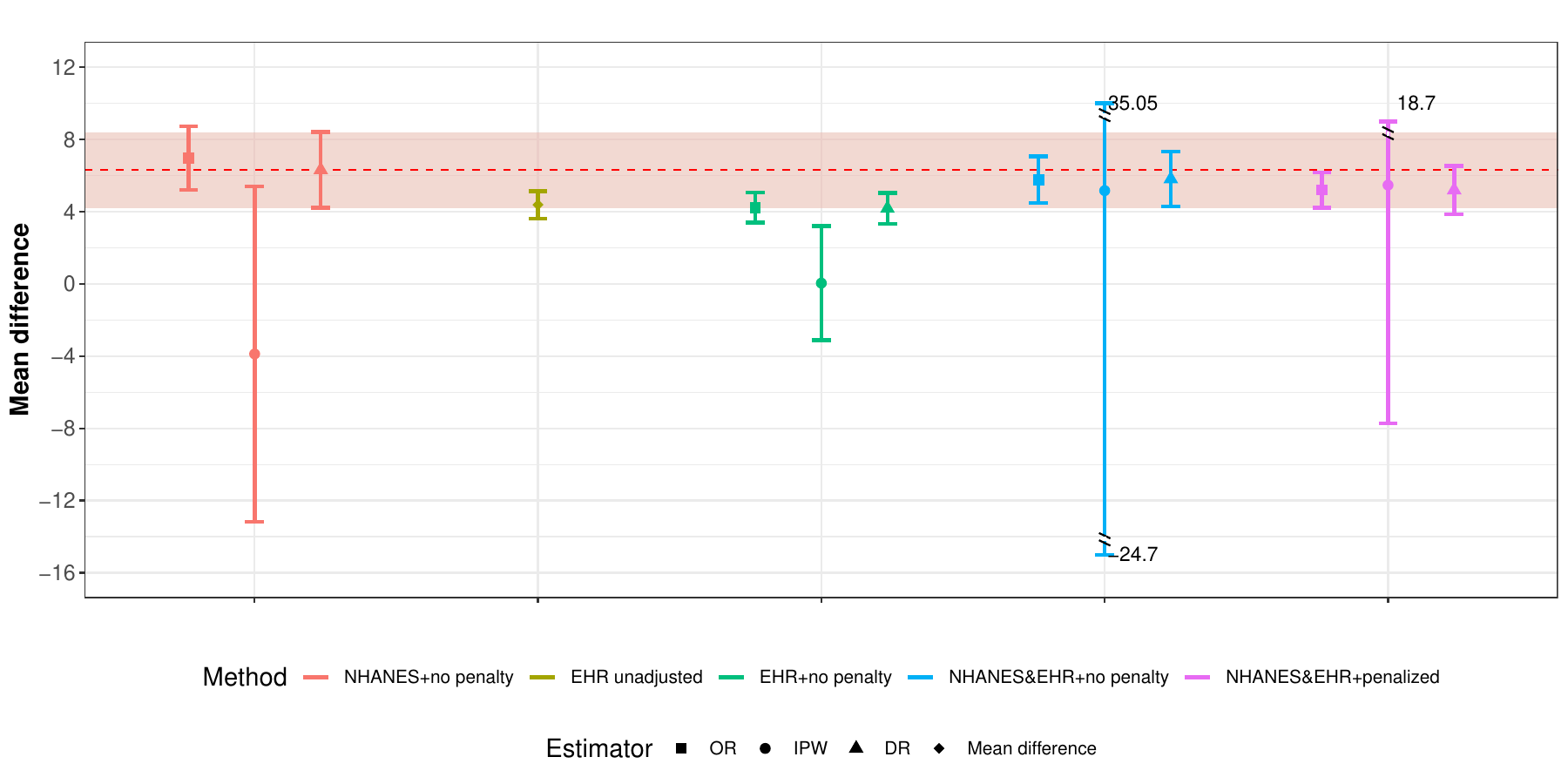}
    \caption{The estimated average treatment effect of severe obesity on systolic blood pressure and 95\% confidence interval (CI). The red horizontal line denotes the point estimate from the doubly robust estimator using exclusively the NHANES data, with the 95\% CI plotted in the banded area. The color of lines represents the method groups based on the data source (NHANES, EHR data from MGI, or both) and the use of penalization (no penalty or penalized). The ``EHR unadjusted'' method represents the sample average of the EHR data that ignores both confounding and selection bias. The shape of the point estimate represents the type of the estimator: outcome regression (OR) estimator, inverse probability weighting (IPW) estimator, and doubly robust (DR) estimator.}
    \label{SBP_ATE_plot}
\end{figure}

\section{Conclusion}

It is challenging to make causal inference of the ATE using non-probability samples due to selection bias and a large number of potential confounders and variables influencing selection. In this paper, we have developed a novel DR estimator of the ATE that integrates a non-probability sample from an EHR database with an external probability sample from designed surveys. To address the problem of having a large number of candidate variables $X$, we have proposed a novel penalized estimating equation procedure that selects variables contributing to confounding relationships and selection mechanisms while simultaneously estimating their effects. 

In practice, the sets of confounders and selection variables are likely to be different. However, with inadequate understanding of the confounding structure and unknown selection mechanism, a safe approach is to start with an inclusive set of variables. Therefore, our proposed method considers a pool of candidate variables $X$ for confounding and selection mechanism together. On the other hand, it is difficult to decide whether $X$ contains all the components for explaining both the confounding structure and selection mechanism, resulting in the possibility of model misspecification. However, our method does not require all models to be correctly specified. This relaxed assumption has a broader application when the underlying models are unknown. 

Our proposed one-step plug-in DR estimator provides an alternative to the two-step post-selection inference approach \citep{yang2020doubly}. There are some fundamental differences which make direct comparison difficult. First, the target estimand is different: While \cite{yang2020doubly} focuses on estimating the population mean, our interest lies in the ATE. This difference leads us to consider a treatment model in addition to the outcome and selection models in \cite{yang2020doubly}'s framework. Second, although it is possible to develop a comparable estimation procedure, we do not pursue this approach because it requires stricter conditions on the variable selection process to ensure all true predictors are selected, as we discuss in Remark 3.

One limitation of our method is that estimating nuisance parameters requires the same set of covariates $X$ for all working models. This problem arises from the constructed estimating equations, which is a common challenge shared with other methods \citep{vermeulen2015bias,yang2020doubly,chen2020doubly}. One remedy is to expand the covariate set until they are matched. For example, with one dimension $X$, if the true outcome model $g(X;\beta)$ is $g(X;\beta) = \beta_0+\beta_1 X+ \beta_2 X^2$ and the true selection model ${\rm logit}\big(\pi_B(X;\alpha) \big)=\alpha_0+\alpha_1 X$, one solution is to expand the selection model to ${\rm logit}\big(\pi_B(X;\alpha) \big)=\alpha_0+\alpha_1 X + \alpha_2 X^2$. Consequently, this condition requires a high quality of the probability sample to have the same set of $X$ as the non-probability sample. Future work should explore the integration of probability and non-probability samples with different sets of covariates.

\section*{Acknowledgment}

The work is supported by R01GM139926.

\section{reference}
\bibliographystyle{apalike}  
\bibliography{references}  
\end{document}


\maketitle

\tableofcontents

\newpage

\beginsupplement

\section{Proof of theorems}

\subsection{Proof of Theorem 1}

We give the proof for the consistency of $\hat{\eta}$ for $\eta_0$ since the proof for the consistency of $\hat{\mu}$ for $\mu_0$ is similar. The proof of Theorem 1 needs the following regularity conditions.

{\itshape

\noindent (a) Interchanging integration and differentiation. Let $\mathcal{B}$ be an open set on $\mathbb{R}^d$. Under the conditions:
(1) $\phi(z;\theta_0, \eta_0, \mu)$ is integrable with respect to the probability measure $F_0(z)$, where $F_0(z)$ represents the true data generating mechanism, (2) for each component in $\mu$, i.e., $\mu_j$, with probability one, the derivative $\partial_{\mu_j} \phi(z;\theta_0, \eta_0, \mu)$ exists, and (3) for all $j$, there is an integrable function $B_j$ such that $|\partial_{\mu_j} \phi(z;\theta_0, \eta_0, \mu) \leq B_j|$ for all $\mu \in \mathcal{B}$, we have for all $j$ that
$$
\partial_{\mu_j} \int \phi (z;\theta_0, \eta_0, \mu) d F_0(z) = \int \partial_{\mu_j} \phi (z;\theta_0, \eta_0, \mu) d F_0(z).
$$

}

Under this condition, the following equality holds for any value of $\mu$.
$$
0  = \partial_{\mu} \big\{ E(\phi (Z;\theta_0, \eta_0, \mu) \big\} = \partial_\mu \int \phi (z;\theta_0, \eta_0, \mu) d F_0(z) = E \big\{ \partial_{\mu} \phi (Z;\theta_0,\eta_0, \mu) \big\}.
$$
In addition, we assume the following conditions also hold for any $\anymu$. For notation simplicity, we omit $Z$ and $\theta_0$ in the $\phi$ and $\phifun$ functions.

{\itshape
\noindent (b) $\sup_{\eta} \left\| \partial_{\mu}\phifun(\eta,\anymu)-E\{\partial_{\mu}\phi(\eta,\anymu)\} \right\| \xrightarrow{p}0$.

\noindent (c) $\inf_{\eta:d(\eta,\eta_0)\geq \epsilon} 
\left\| E\{\partial_{\mu}\phi(\eta,\anymu)\} \right\| \geq 0 = \left\| E\{\partial_{\mu}\phi(\eta_0,\anymu) \} \right\|.
$
}

\begin{proof}

By regularity condition (c), $\forall \epsilon > 0$, there exists a $\delta > 0$, s.t., 
\[
\| E\{\partial_{\mu} \phi(\eta,\tilde{\mu})\} \| - \| E\{\partial_{\mu}\phi(\eta_0,\tilde{\mu})\}\| > \delta,
\]
for all $\eta \in d(\eta,\eta_0)\geq \epsilon$. Furthermore,
\[
\| E\{\partial_{\mu}\phi(\eta,\tilde{\mu})\} - E\{\partial_{\mu}\phi(\eta_0,\tilde{\mu})\} \| \geq 
\| E\{\partial_{\mu} \phi(\eta,\tilde{\mu})\} \| - \| E\{\partial_{\mu}\phi(\eta_0,\tilde{\mu})\}\| > \delta.
\]
Thus, the event $\{d(\hat{\eta},\eta_0)\geq \epsilon)\}$ is contained in the event $\{ \| E\{\partial_{\mu}\phi(\hat{\eta},\tilde{\mu})\} - E\{\partial_{\mu}\phi(\eta_0,\tilde{\mu})\} \| > \delta \}$.

That is,
\[
P(\| \hat{\eta} - \eta_0 \| \geq \epsilon) \leq P(\| E\{\partial_{\mu}\phi(\hat{\eta},\tilde{\mu})\} - E\{\partial_{\mu}\phi(\eta_0,\tilde{\mu})\} \| > \delta).
\]

In addition,
\begin{align*}
    & \| E\{\partial_{\mu}\phi(\hat{\eta},\tilde{\mu})\} - E\{\partial_{\mu}\phi(\eta_0,\tilde{\mu})\} \| \\
    = &      \|E\{\partial_{\mu}\phi(\hat{\eta},\tilde{\mu})\} - \partial_{\mu}\phifun(\hat{\eta},\tilde{\mu}) + \partial_{\mu}\phifun(\hat{\eta},\tilde{\mu})-
    E\{\partial_{\mu}\phi(\eta_0,\tilde{\mu})\} \|\\
    \leq &      \|E\{\partial_{\mu}\phi(\hat{\eta},\tilde{\mu})\} - \partial_{\mu}\phifun(\hat{\eta},\tilde{\mu}) \| + \| \partial_{\mu}\phifun(\hat{\eta},\tilde{\mu})-
    E\{\partial_{\mu}\phi(\eta_0,\tilde{\mu})\} \| \\
    \leq &      \sup_{\eta }\|E\{\partial_{\mu}\phi(\eta,\tilde{\mu})\} - \partial_{\mu}\phifun(\eta,\tilde{\mu}) \| + \| \partial_{\mu}\phifun(\hat{\eta},\tilde{\mu})-
    E\{\partial_{\mu}\phi(\eta_0,\tilde{\mu})\} \|.
\end{align*}
By regularity condition (b), 
\[
\sup_{\eta }\|E\{\partial_{\mu}\phi(\eta,\tilde{\mu})\} - \partial_{\mu}\phifun(\eta,\tilde{\mu}) \| \xrightarrow{p} 0.
\]
\noindent For the second term, by the property of $\hat{\eta}$, $\| \partial_{\mu}\phifun(\hat{\eta},\tilde{\mu})\| =o_p(1)$. It follows that
\[
\| \partial_{\mu}\phifun(\hat{\eta},\tilde{\mu})- E\{\partial_{\mu}\phi(\eta_0,\tilde{\mu})\} \| = \| \partial_{\mu}\phifun(\hat{\eta},\tilde{\mu}) \| = o_p(1).
\]

Therefore,
\[
\lim_{n \to \infty} P(\| E\{\partial_{\mu}\phi(\hat{\eta},\tilde{\mu})\} - E\{\partial_{\mu}\phi(\eta_0,\tilde{\mu})\} \| > \delta) = 0, \text{ so that }
\]
\[
\lim_{n \to \infty} P(\| \hat{\eta} - \eta_0 \| \geq \epsilon) = 0.
\]
\end{proof}

\subsection{Proof of Theorem 2}

\begin{proof}

To prove Theorem 2(c), we construct $\Tilde{\omega}$ in a way such that $\Tilde{\omega}_{\mathcal{M}_\omega} = \omega_{\mathcal{M}_\omega}^{\ast} - \Omega_{\mathcal{M}_\omega,\mathcal{M}_\omega}^{-1} \Ufun_{\mathcal{M}_\omega} (\omega^{\ast})$ and for $j \in \mathcal{M}_{\omega}^c$, $\Tilde{\omega}_j = 0$, where $\Ufun_{\mathcal{M}_\omega}(\omega^\ast)$ denotes a subvector of $\Ufun(\omega^\ast)$ with indexes belonging to $\mathcal{M}_\omega$ and $\Omega_{\mathcal{M}_\omega,\mathcal{M}_\omega}$ denotes a submatrix of $\Omega$ with row and column indexes belonging to $\mathcal{M}_\omega$. By Assumption 3(a), $\| \Ufun_{\mathcal{M}_\omega} (\omega^{\ast})\| = O_p(n^{-1/2})$. Since $\Omega_{\mathcal{M}_\omega,\mathcal{M}_\omega}$ is non-singular, $ \| \Omega_{\mathcal{M}_\omega,\mathcal{M}_\omega}^{-1} \| = \sigma_{\min} (\Omega_{\mathcal{M}_\omega,\mathcal{M}_\omega}) = O_p(1)$, where $\sigma_{\min}$ denotes the smallest singular value. Therefore, 
\begin{equation*}
    \| \Tilde{\omega} - \omega^{\ast}\| = \| \Omega_{\mathcal{M}_\omega,\mathcal{M}_\omega}^{-1} \Ufun_{\mathcal{M}_\omega} \| \leq  \| \Omega_{\mathcal{M}_\omega,\mathcal{M}_\omega}^{-1} \| \cdot \|\Ufun_{\mathcal{M}_\omega} \| = O_p(n^{-1/2}).
\end{equation*}
Next, we will show that $\Tilde{\omega}$ we construct is an approximate zero-crossing for $\Ufun^p(\omega)$. For $j \in \mathcal{M}_\omega$, 
\begin{align}
    & \sqrt{n} \Ufun_j^p(\Tilde{\omega} + \epsilon e_j) = \sqrt{n} \big\{ \Ufun_j(\Tilde{\omega} + \epsilon e_j) + q_{\lambda_\omega} (| \Tilde{\omega}_j + \epsilon |)sgn(\Tilde{\omega}_j + \epsilon) \big\} \nonumber\\
    & = \sqrt{n} \Ufun_j (\Tilde{\omega} + \epsilon e_j) + o(1) \label{thm2_eq1} \\ 
    & = \sqrt{n} \Ufun_j(\omega^{\ast}) + \sqrt{n} \big[ \Omega_{\mathcal{M}_\omega,\mathcal{M}_\omega} \big\{- \Omega_{\mathcal{M}_\omega,\mathcal{M}_\omega}^{-1} \Ufun_{\mathcal{M}_\omega} (\omega^{\ast}) \big\} \big]_j + \sqrt{n} \Omega_{j,j} \epsilon  + o(1) \label{thm2_eq2}\\ 
    & = \sqrt{n} \Ufun_j(\omega^{\ast}) - \sqrt{n} \Ufun_j(\omega^{\ast}) + \sqrt{n} \Omega_{j,j} \epsilon + o(1) \nonumber \\
    & = \sqrt{n} \Omega_{j,j} \epsilon + o(1), \label{thm2_j_M}
\end{align}
\noindent where $[\cdot]_j$ in \eqref{thm2_eq2} denotes the element within the bracket correponding to index $j$, and $\Omega_{j,j}$ is the diagnal element in $\Omega$ corresponding to index $j$. Equation \eqref{thm2_eq1} is true under the Assumption 3(b) and \eqref{thm2_eq2} is the Taylor series under the Assumption 3(a). Similarly,
\begin{equation*}
    \sqrt{n} \Ufun_j^p(\Tilde{\omega} - \epsilon e_j) = -\sqrt{n} \Omega_{j,j} \epsilon + o(1).
\end{equation*}
\noindent Consequently, we have $n \Ufun_j^p (\Tilde{\omega} + \epsilon e_j) \Ufun_j^p(\Tilde{\omega} - \epsilon e_j) = -n\Omega_{j,j}^2\epsilon^2+o(1)$ so that 
\begin{equation*}
        \uplim_{N\rightarrow\infty} P\big\{ \uplim_{\epsilon \rightarrow 0+}  n \Ufun_j^p(\Tilde{\omega} + \epsilon e_j) \Ufun_j^p(\Tilde{\omega} - \epsilon e_j) \leq 0 \big\}  = 1.
\end{equation*}

For $j \in \mathcal{M}_{\omega}^c$,
\begin{align}
    & \sqrt{n} \Ufun_j^p(\Tilde{\omega} + \epsilon e_j) = \sqrt{n} \big\{\Ufun_j(\Tilde{\omega} + \epsilon e_j) + q_{\lambda_\omega} (\epsilon) \big\} \nonumber\\
    & = \sqrt{n} \Ufun_j(\Tilde{\omega} + \epsilon e_j) + \sqrt{n}q_{\lambda_\omega} (\epsilon)  \nonumber \\ 
    & = \sqrt{n} \Ufun_j(\omega^{\ast}) + \sqrt{n} \big[\Omega(\Tilde{\omega}-\omega^{\ast}) \big]_j + \sqrt{n} \Omega_{j,j} \epsilon + \sqrt{n}q_{\lambda_\omega} (\epsilon) \nonumber\\ 
    & = \sqrt{n} \Ufun_j(\omega^{\ast}) +  0     + \sqrt{n} \Omega_{j,j} \epsilon + \sqrt{n}q_{\lambda_\omega}(\epsilon). \label{thm2_j_Mc}
\end{align}
Denote $\sqrt{n} \Ufun_j(\omega^{\ast})     + \sqrt{n} \Omega_{j,j} \epsilon$ as $C_n$. Under the Assumption 3(c), $\sqrt{n} \Ufun_j(\omega^{\ast}) = O_p(1)$ and we choose $\epsilon$ such at $\sqrt{n} \Omega_{j,j}\epsilon = o(1)$. Therefore,  $\sqrt{n} \Ufun_j^p(\Tilde{\omega} + \epsilon e_j) = C_n + \sqrt{n}q_{\lambda_\omega} (\epsilon) $, where $C_n = O_p(1)$. Similarly, we have $\sqrt{n} \Ufun_j^p(\Tilde{\omega} - \epsilon e_j) =  C_n - \sqrt{n}q_{\lambda_\omega} (\epsilon)$, where $C_n = O_p(1)$.

Since 
\begin{equation*}
    n \Ufun_j^p(\Tilde{\omega} + \epsilon e_j) \Ufun_j^p(\Tilde{\omega} - \epsilon e_j) = -n q_{\lambda_\omega}^2 (\epsilon) + C_n^2
\end{equation*}

and under the Assumption 3(c) that $\sqrt{n}q_{\lambda_\omega}(\epsilon) \rightarrow \infty$, $-n q_{\lambda_\omega}^2 (\epsilon)$ is dominant over $C_n^2$. We have:
\begin{equation*}
        \lim_{N\rightarrow\infty} P\big\{ \uplim_{\epsilon \rightarrow 0+}  n \Ufun_j^p(\Tilde{\omega} + \epsilon e_j) \Ufun_j^p(\Tilde{\omega} - \epsilon e_j) \leq 0 \big\} = 
        \lim_{N\rightarrow\infty} P\bigg\{ \uplim_{\epsilon \rightarrow 0+} \big( - n q_{\lambda_\omega}^2 (\epsilon) + C_n^2 \big) \leq 0 \bigg\} = 1.
\end{equation*}

Therefore, $\Tilde{\omega}$ is an approximate root-n-consistent zero-crossing of $\Ufun^p(\omega)$. Theorem 2(b) is true since $\Tilde{\omega}_j = 0$ for all $j \in \mathcal{M}_\omega^c$ by our construction. Next we would like to show that Theorem 2(a) is true. 
For $j \in \mathcal{M}_\omega$, 
\begin{align}
    &   \Ufun_j^p(\Tilde{\omega}) =   \big\{ \Ufun_j(\Tilde{\omega}) + q_{\lambda_\omega} (| \Tilde{\omega}_j |)sgn(\Tilde{\omega}_j ) \big\} \nonumber\\
    & =   \Ufun_j (\Tilde{\omega}) + o(1) \label{thm2a_eq1} \\ 
    & =   \Ufun_j(\omega^{\ast}) +   \big[ \Omega_{\mathcal{M}_\omega,\mathcal{M}_\omega} \big\{- \Omega_{\mathcal{M}_\omega,\mathcal{M}_\omega}^{-1} \Ufun_{\mathcal{M}_\omega} (\omega^{\ast}) \big\} \big]_j  + o(1) \label{thm2a_eq2}\\ 
    & =   \Ufun_j(\omega^{\ast}) -   \Ufun_j(\omega^{\ast}) + o(1) \nonumber \\
    & = o(1), \label{thm2a_j_M}
\end{align}
Equation \eqref{thm2a_eq1} is true under the Assumption 3(b) and \eqref{thm2a_eq2} is the Taylor series under the Assumption 3(a). For $j \in \mathcal{M}_{\omega}^c$,
\begin{align}
    &   \Ufun_j^p(\Tilde{\omega}) =   \big\{\Ufun_j(\Tilde{\omega} ) + q_{\lambda_\omega} (0)sgn(0) \big\} \nonumber\\
    & =   \Ufun_j(\Tilde{\omega}) \nonumber \\ 
    & =   \Ufun_j(\omega^{\ast}) +   [\Omega(\Tilde{\omega}-\omega^{\ast})]_j \label{thm2a_eq3}\\ 
    & =   \Ufun_j(\omega^{\ast}) \nonumber \\
    & = o_p(1). \label{thm2a_j_Mc}
\end{align}
Note that \eqref{thm2a_eq3} is the Taylor series under the Assumption 3(a). Under Assumption 3(a), \eqref{thm2a_j_Mc} is true. Therefore, combining \eqref{thm2a_j_M} and \eqref{thm2a_j_Mc}, Theorem 2(a) is true. 

\end{proof}

\section{Asymptotic variance of the DR estimator}

\subsection{Derivation of $V_1$ and $V_2$}

The first term $V_1$ is the sampling variance of the Horvitz-Thompson estimator for the probability sample $A$. Since $I_{A,i} \ind I_{A,j}|X_i,X_j$,
\begin{align}
        V_1 &= n E_{I_B,T,X,Y}\Big[ V_{I_A|I_B,T,X,Y} \Big\{ N^{-1} \sum_{i=1}^N I_{A,i}d_{A,i} \big( \EYT(X_i; \beta^\ast) - \EYC(X_i; \gamma^\ast) \big) \Big\} \Big] \nonumber \\
        &= E_{X}\Big[ n/N^2 \sum_{i=1}^{N} V_{I_A|X} \Big\{ I_{A,i}d_{A,i} \big( \EYT(X_i; \beta^\ast) - \EYC(X_i; \gamma^\ast) \big) \Big\} \Big] \nonumber \\
        &= E_{X} \Big\{ n/N^2 \sum_{i=1}^{N}   (d_{A,i} - 1)\big( \EYT(X_i; \beta^\ast) - \EYC(X_i; \gamma^\ast) \big)^2 \Big\} \nonumber. 
\end{align}
For the second term $V_2$, note that $E_{I_A|I_B,T,X,Y} \big[ \hat{\theta}_{\rm DR}(\omega^\ast) - \theta_0 \big]$ can be expressed as
\begin{align*}
    N^{-1} \sum_{i=1}^N \Big\{ & \EYT(X_i; \beta^\ast) +  \frac{I_{B,i}T_i}{\pi_B(X_i;\alpha^{\ast}) \pi_T(X_i; \tau^{\ast})} \big( Y_i(1) - \EYT(X_i;\beta^{\ast}) \big) - \\
    & \EYC(X_i;\gamma\ast) -  \frac{I_{B,i} (1-T_i )}{\pi_B(X_i;\alpha^{\ast}) (1-\pi_T(X_i; \tau^{\ast}) )} \big( Y_i(0) - \EYC(X_i;\gamma^{\ast}) \big)  - \theta_0 \Big\}.
\end{align*}
Thus,
\begin{align*}
    V_2 =  n E_{I_B, T, X, Y} \bigg[ \bigg\{ E_{I_A| I_B, T,  X, Y} & \big(  \hat{\theta}_{\rm DR}(\omega^\ast) - \theta_0) \big) \bigg\}^2 \bigg] \\
    = E_{I_B, T, X, Y} \bigg[ n/N^2 \sum_{i=1}^N \bigg\{ & \EYT(X_i; \beta^\ast) +  \frac{I_{B,i}T_i}{\pi_B(X_i;\alpha^{\ast}) \pi_T(X_i; \tau^{\ast})} \big( Y_i(1) - \EYT(X_i;\beta^{\ast}) \big) - \\
    & \EYC(X_i;\gamma^\ast) -  \frac{I_{B,i} (1-T_i )}{\pi_B(X_i;\alpha^{\ast}) (1-\pi_T(X_i; \tau^{\ast}) )} \big( Y_i(0) - \EYC(X_i;\gamma^{\ast}) \big)  - \theta_0  \bigg\}^2     \bigg] + \\
    E_{I_B, T, X, Y} \bigg[ n/N^2 \sum_{i=1}^N \sum_{j\neq i} 
     \bigg\{  \bigg( & \EYT(X_i; \beta^\ast) +  \frac{I_{B,i}T_i}{\pi_B(X_i;\alpha^{\ast}) \pi_T(X_i; \tau^{\ast})} \big( Y_i(1) - \EYT(X_i;\beta^{\ast}) \big) - \\
     & \EYC(X_i;\gamma^\ast) -  \frac{I_{B,i} (1-T_i )}{\pi_B(X_i;\alpha^{\ast}) (1-\pi_T(X_i; \tau^{\ast}) )} \big( Y_i(0) - \EYC(X_i;\gamma^{\ast}) \big)  - \theta_0 \bigg)  \bigg\} \ \cdot \\
     \bigg\{  \bigg( & \EYT(X_j; \beta^\ast) +  \frac{I_{B,j}T_j}{\pi_B(X_j;\alpha^{\ast}) \pi_T(X_j; \tau^{\ast})} \big( Y_j(1) - \EYT(X_j;\beta^{\ast}) \big) - \\
     & \EYC(X_j;\gamma^\ast) -  \frac{I_{B,j} (1-T_j )}{\pi_B(X_j;\alpha^{\ast}) (1-\pi_T(X_j; \tau^{\ast}) )} \big( Y_j(0) - \EYC(X_j;\gamma^{\ast}) \big)  - \theta_0 \bigg)  \bigg\} \bigg] .
\end{align*}

Since $T_i \ind T_j|X_i,X_j,I_{B,i} = I_{B,j}=1; I_{B,i} \ind I_{B,j}|X_i, X_j; \big( Y_i(0), Y_i(1) \big) \ind \big\{ Y_j(0), Y_j(1) \big\}|X_i, X_j$, the second term in $V_2$ is 0 under the double robustness condition. Therefore,
\begin{align*}
    V_2 = \frac{n}{N} E_{I_B, T, X, Y} \bigg[
    \bigg\{ & \frac{I_{B,i}T_i}{\pi_B(X_i;\alpha^{\ast}) \pi_T(X_i; \tau^{\ast})} \big( Y_i(1) - \EYT(X_i;\beta^{\ast}) \big) -  \\
    & \frac{I_{B,i} (1-T_i )}{\pi_B(X_i;\alpha^{\ast}) (1-\pi_T(X_i; \tau^{\ast}) )} \big( Y_i(0) - \EYC(X_i;\gamma^{\ast}) \big) +  \big( \EYT(X_i; \beta^\ast) - \EYC(X_i;\gamma^\ast) - \theta_0 \big)  \bigg\}^2 \bigg].
\end{align*}

\subsection{Estimator for $V_1$}

The design-consistent estimator is commonly used in survey sampling literature \cite{kott1990estimating, deville1992calibration, deville1994variance, breidt2000local}. 
Different from the classical large-sample framework, the source of randomness in survey samples is the "design" itself, such as the sampling mechanism, and people often consider other quantities, such as $X$ and $Y$, are fixed \cite{kott1990estimating, deville1992calibration, deville1994variance, breidt2000local}. Specifically, $\hat{V}_1$ is a design-consistent estimator for $V_1$ in the sense that
$$
\lim_{N\rightarrow \infty} E_{I_A|X} \big\{\hat{V}_1 - V_1 \big\} = 0.
$$
To prove this, we assume that the following regularity conditions hold: (a) $\lim\limits_{N \rightarrow \infty} N^{-1} \sum_{i=1}^N \big\{ \EYT(X_i;\beta^\ast) - \EYC(X_i;\gamma^\ast) \big\}^4 <\infty$; and (b) $max_{i} (d_{A,i}-1)^3 <\infty$. It is equivalent to showing that 
$$
\lim\limits_{N\rightarrow \infty} E_{I_A|X} \big\{(\hat{V}_1 - V_1)^2 \big\} = 0.
$$
Since 
\begin{align*}
    & E_{I_A|X} \big\{(\hat{V}_1 - V_1)^2 \big\} \\
    &= \frac{n^2}{N^4} \sum_{i=1}^N E_{I_A|X}\big\{(I_{A,i}d_{A,i} - 1)^2\big\} (d_{A,i}-1)^2 \big\{ \EYT(X_i;\beta^\ast) - \EYC(X_i;\gamma^\ast) \big\}^4 \\
    &= \frac{n^2}{N^4} \sum_{i=1}^N E_{I_A|X}\big\{I_{A,i}d_{A,i}^2 -2 I_{A,i}d_{A,i} + 1\big\} (d_{A,i}-1)^2 \big\{ \EYT(X_i;\beta^\ast) - \EYC(X_i;\gamma^\ast) \big\}^4 \\
    &= \frac{n^2}{N^4} \sum_{i=1}^N (d_{A,i}-1)^3 \big\{ \EYT(X_i;\beta^\ast) - \EYC(X_i;\gamma^\ast) \big\}^4,
\end{align*}
$\lim\limits_{N\rightarrow \infty} E_{I_A|X} \big\{(\hat{V}_1 - V_1)^2 \big\} = 0$ under the above regularity conditions and Assumption 2.

\subsection{Estimator of $V_2$}

We expand the squared term in $V_2$ and obtain 
\begin{align*}
    V_2 &= \frac{n}{N} E \Bigg[ \bigg\{ \frac{I_{B,i}T_i}{\pi_{B}(X_i; \alpha^{\ast} ) \pi_{T}(X_i;\tau^{\ast})}  \bigg( Y_i(1) - \EYT(X_i;\beta^{\ast}) \bigg) \bigg\}^2 \Bigg] \\ 
    & + \frac{n}{N} E \Bigg[ \bigg\{ \frac{I_{B,i}(1-T_i )}{\pi_{B}(X_i; \alpha^{\ast} ) (1-\pi_{T}(X_i;\tau^{\ast}) )} \bigg( Y_i(0) - \EYC(X_i;\gamma^{\ast}) \bigg) \bigg\}^2 \Bigg] \\ 
    & + \frac{n}{N} E \Bigg\{ \bigg(\EYT(X_i;\beta^\ast) - \EYC(X_i;\gamma^\ast) - \theta_0 \bigg)^2 \Bigg\} \\ 
    & - \frac{2n}{N} E \Bigg\{ \frac{I_{B,i}T_i}{\pi_{B}(X_i; \alpha^{\ast} ) \pi_{T}(X_i;\tau^{\ast})}  \bigg( Y_i(1) - \EYT(X_i;\beta^{\ast}) \bigg) \cdot \frac{I_{B,i}(1-T_i )}{\pi_{B}(X_i; \alpha^{\ast} ) (1-\pi_{T}(X_i;\tau^{\ast}) )} \bigg( Y_i(0) - \EYC(X_i;\gamma^{\ast}) \bigg) \Bigg\} \\ 
    & + \frac{2n}{N} E \Bigg\{ \frac{I_{B,i}T_i}{\pi_{B}(X_i; \alpha^{\ast} ) \pi_{T}(X_i;\tau^{\ast})}  \bigg( Y_i(1) - \EYT(X_i;\beta^{\ast}) \bigg) \cdot \bigg(\EYT(X_i;\beta^\ast) - \EYC(X_i;\gamma^\ast) - \theta_0 \bigg)  \Bigg\} \\ 
    & - \frac{2n}{N} E \Bigg\{\frac{I_{B,i}(1-T_i )}{\pi_{B}(X_i; \alpha^{\ast} ) (1-\pi_{T}(X_i;\tau^{\ast}) )} \bigg( Y_i(0) - \EYC(X_i;\gamma^{\ast}) \bigg) \cdot \bigg(\EYT(X_i;\beta^\ast) - \EYC(X_i;\gamma^\ast) - \theta_0 \bigg) \Bigg\} \\ 
    & := S_1 + S_2 + S_3 + S_4 + S_5 + S_6,
\end{align*}
\noindent where the expectation is taken with respect to $I_B, T, X, Y$ and $S_4=0$. The remaining terms are consistently estimated as:
\begin{align*}
    \hat{S}_1 &= \frac{n}{N^2} \sum_{i=1}^N \bigg\{ \frac{I_{B,i}T_i}{\pi_{B}(X_i; \hat{\alpha}^p ) \pi_{T}(X_i;\hat{\tau}^p)}  \bigg( Y_i - \EYT(X_i;\hat{\beta}^p) \bigg) \bigg\}^2, \\
    \hat{S}_2 &= \frac{n}{N^2} \sum_{i=1}^N \bigg\{ \frac{I_{B,i}(1-T_i )}{\pi_{B}(X_i; \hat{\alpha}^p ) (1-\pi_{T}(X_i;\hat{\tau}^p) )} \bigg( Y_i - \EYC(X_i;\hat{\gamma}^p) \bigg) \bigg\}^2, \\
    \hat{S}_3 &= \frac{n}{N^2} \sum_{i=1}^N  I_{A,i}d_{A,i}\bigg(\EYT(X_i;\hat{\beta}^p) - \EYC(X_i;\hat{\gamma}^p) - \hat{\theta}_{\rm DR}(\pSol) \bigg)^2, \\
    \hat{S}_5 &= \frac{2n}{N^2} \sum_{i=1}^N \frac{I_{B,i}T_i}{\pi_{B}(X_i; \hat{\alpha}^p ) \pi_{T}(X_i;\hat{\tau}^p)}  \bigg( Y_i - \EYT(X_i;\hat{\beta}^p) \bigg) \cdot \bigg(\EYT(X_i;\hat{\beta}^p) - \EYC(X_i;\hat{\gamma}^p) - \hat{\theta}_{\rm DR}(\pSol) \bigg), \\
    \hat{S}_6 &= -\frac{2n}{N^2} \sum_{i=1}^N \frac{I_{B,i}(1-T_i )}{\pi_{B}(X_i; \hat{\alpha}^p ) (1-\pi_{T}(X_i;\hat{\tau}^p) )} \bigg( Y_i - \EYC(X_i;\hat{\gamma}^p) \bigg) \cdot \bigg(\EYT(X_i;\hat{\beta}^p) - \EYC(X_i;\hat{\gamma}^p) - \hat{\theta}_{\rm DR}(\pSol) \bigg).
\end{align*}

\noindent Therefore, $V_2$ is consistently estimated as $\hat{S}_1 + \hat{S}_2+ \hat{S}_3+ \hat{S}_5 + \hat{S}_6$.

\section{Variance estimators in the data example}
\subsection{Variance estimator for unpenalized estimating equations}

Let $\sum_{i}^N \psi(Z_i;\theta,\omega) $ be the estimating function for $\theta$, the average treatment effect, and $\omega$, the nuisance parameters. $(\hat{\theta}, \hat{\omega})$ jointly solves 
\begin{equation*}
    \sum_{i=1}^N \psi(Z_i;\hat{\theta},\hat{\omega}) = 0.
\end{equation*}
Denote $\hat{\theta} \xrightarrow{p} \theta^\ast$ and $\hat{\omega} \xrightarrow{p} \omega^\ast$. By Taylor series expansion, we have
\begin{equation*}
    0 = N^{-1/2} \sum_{i=1}^N \psi (Z_i; \theta^\ast,\omega^\ast) + N^{-1/2} \sum_{i=1}^N \frac{\partial \psi (Z_i; \theta^\ast,\omega^\ast)}{\partial (\theta^{\top}, \omega^{\top})} 
    \begin{pmatrix}
        \hat{\theta} - \theta^{\ast} \\
        \hat{\omega} - \omega^{\ast} 
    \end{pmatrix}  + o_p(1).
\end{equation*}
It follows that 
\begin{equation*}
    \sqrt{N}
        \begin{pmatrix}
        \hat{\theta} - \theta^{\ast} \\
        \hat{\omega} - \omega^{\ast} 
    \end{pmatrix} = \Big\{ N^{-1} \sum_{i=1}^N \frac{\partial \psi (Z_i; \theta^\ast,\omega^\ast)}{\partial (\theta^{\top}, \omega^{\top})} \Big\}^{-1} \Big\{ N^{-1/2} \sum_{i=1}^N \psi (Z_i; \theta^\ast,\omega^\ast) \Big\} + o_p(1).
\end{equation*}
Therefore, 
\begin{equation*}
    \sqrt{N}
    \begin{pmatrix}
        \hat{\theta} - \theta^{\ast} \\
        \hat{\omega} - \omega^{\ast} 
    \end{pmatrix} 
    \xrightarrow{d} MVN(0, A^{-1}_{\theta^\ast, \omega^\ast}  B_{\theta^\ast, \omega^\ast} \{A^{-1}_{\theta^\ast, \omega^\ast}\}^{\top} ),
\end{equation*}
\noindent where 
\begin{equation*}
    A_{\theta^\ast, \omega^\ast} = E\Big\{ \frac{\partial \psi (Z_i; \theta^\ast,\omega^\ast)}{\partial (\theta^{\top}, \omega^{\top})} \Big\}, B_{\theta^\ast, \omega^\ast} = E\big\{ \psi (Z_i; \theta^\ast,\omega^\ast) \psi^{\top} (Z_i; \theta^\ast,\omega^\ast) \big\}.
\end{equation*}
The variance estimator $\widehat{Var} 
    \begin{pmatrix}
        \hat{\theta} \\
        \hat{\omega} 
    \end{pmatrix}$ for $(\hat{\theta},\hat{\omega})$ is:
\begin{equation*}
    N^{-1}\Big\{ \Big(N^{-1}\sum_{i=1}^N \frac{\partial \psi (Z_i; \hat{\theta},\hat{\omega})}{\partial (\theta^{\top}, \omega^{\top})} \Big)^{-1}\Big\} \cdot
    \Big\{ \Big(N^{-1}\sum_{i=1}^N \psi (Z_i; \hat{\theta},\hat{\omega}) \psi^{\top} (Z_i; \hat{\theta},\hat{\omega}) \Big)\Big\} \cdot
    \Big\{ \Big(N^{-1}\sum_{i=1}^N \frac{\partial \psi (Z_i; \hat{\theta},\hat{\omega})}{\partial (\theta^{\top}, \omega^{\top})} \Big)^{-1}\Big\} ^{\top}.
\end{equation*}

We now present the specific form the $\psi(\cdot)$ function for different estimators.

\noindent \textit{Outcome regression (OR) estimator using the NHANES data only:}
\begin{equation*}
    \sum_{i=1}^N \psi(Z_i; \theta, \omega) = \sum_{i=1}^N
    \begin{pmatrix}
        \theta - I_{A,i}d_{A,i} \big(\Tilde{\EYT}(X_i;\beta) - \Tilde{\EYC}(X_i; \gamma) \big) \\
        I_{A,i}T_i\big(Y_i-\Tilde{\EYT}(X_i;\beta)\big)X_i \\
        I_{A,i}(1-T_i)\big(Y_i-\Tilde{\EYC}(X_i;\gamma)\big)X_i
    \end{pmatrix},
\end{equation*}
where $\omega = (\beta,\gamma)$.

\noindent  \textit{Inverse probability weighting (IPW) estimator using the NHANES data only:}
\begin{equation*}
    \sum_{i=1}^N \psi(Z_i; \theta, \omega) = \sum_{i=1}^N
    \begin{pmatrix}
        \theta -  I_{A,i}d_{A,i} \big( \frac{T_i}{\Tilde{\pi}_T(X_i;\tau)}Y_i - \frac{1-T_i}{1-\Tilde{\pi}_T(X_i;\tau)}Y_i \big) \\
        I_{A,i} \big(T_i-\Tilde{\pi}_T(X_i;\tau)\big)X_i
    \end{pmatrix},
\end{equation*}
where $\omega = \tau$.

\noindent  \textit{Double robust (DR) estimator using the NHANES data only:}
\begin{equation*}
    \sum_{i=1}^N \psi(Z_i; \theta, \omega) = \sum_{i=1}^N
    \begin{pmatrix}
        \theta - I_{A,i} d_{A,i} \Tilde{o}_i\\
        I_{A,i}T_i\big(Y_i-\Tilde{\EYT}(X_i;\beta)\big)X_i \\
        I_{A,i}(1-T_i)\big(Y_i-\Tilde{\EYC}(X_i;\gamma)\big)X_i \\
        I_{A,i} \big(T_i-\Tilde{\pi}_T(X_i;\tau)\big)X_i
    \end{pmatrix},
\end{equation*}
where $\omega = (\beta,\gamma, \tau)$, and $\Tilde{o_i} = \Big\{ \Tilde{\EYT}(X_i;\beta) + \frac{T_i}{\Tilde{\pi}_T(X_i;\tau)} \big(Y_i-\Tilde{\EYT}(X_i;\beta) \big)   \Big\} - \Big\{ \Tilde{\EYC}(X_i;\gamma) + \frac{1-T_i}{1-\Tilde{\pi}_T(X_i;\tau)} \big(Y_i-\Tilde{\EYC}(X_i;\gamma) \big)   \Big\}$.

\noindent  \textit{OR estimator using EHR data only:}
\begin{equation*}
    \sum_{i=1}^{n_B} \psi(Z_i; \theta, \omega) = \sum_{i=1}^{n_B}
    \begin{pmatrix}
        \theta - \big(\EYT(X_i;\beta) - \EYC(X_i; \gamma) \big) \\
        T_i\big(Y_i-\EYT(X_i;\beta)\big)X_i \\
        (1-T_i)\big(Y_i-\EYC(X_i;\gamma)\big)X_i
    \end{pmatrix},
\end{equation*}
where $\omega = (\beta,\gamma)$.

\noindent  \textit{IPW estimator using EHR data only:}
\begin{equation*}
    \sum_{i=1}^{n_B} \psi(Z_i; \theta, \omega) = \sum_{i=1}^{n_B}
    \begin{pmatrix}
        \theta -  \big( \frac{T_i}{\pi_T(X_i;\tau)}Y_i - \frac{1-T_i}{1-\pi_T(X_i;\tau)}Y_i \big) \\
        \big(T_i-\pi_T(X_i;\tau)\big)X_i
    \end{pmatrix},
\end{equation*}
where $\omega = \tau$.

\noindent  \textit{DR estimator using EHR data only:}
\begin{equation*}
    \sum_{i=1}^{n_B} \psi(Z_i; \theta, \omega) = \sum_{i=1}^{n_B}
    \begin{pmatrix}
        \theta - o_i \\
        T_i\big(Y_i-\EYT(X_i;\beta)\big)X_i \\
        (1-T_i)\big(Y_i-\EYC(X_i;\gamma)\big)X_i \\
        \big(T_i-\pi_T(X_i;\tau)\big)X_i
    \end{pmatrix},
\end{equation*}
where $\omega = (\beta,\gamma, \tau)$ and $o_i = \Big\{ \EYT(X_i;\beta) + \frac{T_i}{\pi_T(X_i;\tau)} \big(Y_i-\EYT(X_i;\beta) \big)   \Big\}- \Big\{ \EYC(X_i;\gamma) + \frac{1-T_i}{1-\pi_T(X_i;\tau)} \big(Y_i-\EYC(X_i;\gamma) \big)   \Big\} $.

\noindent  \textit{OR estimator using the NHANES and EHR data:}
\begin{equation*}
    \sum_{i=1}^{N} \psi(Z_i; \theta, \omega) = \sum_{i=1}^{N}
    \begin{pmatrix}
        \theta - I_{A,i}d_{A,i} \big(\EYT(X_i;\beta) - \EYC(X_i; \gamma) \big) \\
        I_{B,i} T_i\big(Y_i-\EYT(X_i;\beta)\big)X_i \\
        I_{B,i} (1-T_i)\big(Y_i-\EYC(X_i;\gamma)\big)X_i
    \end{pmatrix},
\end{equation*}
where $\omega = (\beta,\gamma)$.

\noindent  \textit{IPW estimator using the NHANES and EHR data:}
\begin{equation*}
    \sum_{i=1}^{N} \psi(Z_i; \theta, \omega) = \sum_{i=1}^{N}
    \begin{pmatrix}
        \theta -  \big( \frac{I_{B,i} T_i}{\pi_{B}(X_i;\alpha)\pi_{T}(X_i;\tau)}Y_i - \frac{I_{B,i}(1-T_i)}{\pi_{B}(X_i;\alpha)\big(1-\pi_{T}(X_i;\tau)\big)}Y_i \big) \\
        (I_{B,i}/\pi_{B}(X_i;\alpha) - I_{A,i} d_{A,i})X_i \\
        I_{B,i}\big(T_i-\pi_T(X_i;\tau)\big)X_i
    \end{pmatrix},
\end{equation*}
where $\omega = (\alpha,\tau)$.

\noindent  \textit{DR estimator using the NHANES and EHR data:}
\begin{equation*}
        \sum_{i=1}^{N} \psi(Z_i; \theta, \omega)
    = \sum_{i=1}^{N}
    \begin{pmatrix}
        \theta -  o_i\\
        (I_{B,i}/\pi_{B}(X_i;\alpha) - I_{A,i} d_{A,i})X_i \\
        I_{B,i}\big(T_i-\pi_T(X_i;\tau)\big)X_i \\
        I_{B,i}T_i\big(Y_i-\EYT(X_i;\beta)\big)X_i \\
        I_{B,i}(1-T_i)\big(Y_i-\EYC(X_i;\gamma)\big)X_i
    \end{pmatrix},
\end{equation*}
where $\omega = (\alpha,\tau,\alpha,\beta)$ and $o_i = \Big\{ I_{A,i}d_{A,i} \EYT(X_i;\beta) + \frac{I_{B,i} T_i}{\pi_{B}(X_i;\alpha)\pi_{T}(X_i;\tau)} \big(Y_i-\EYT(X_i;\beta) \big)   \Big\} - \Big\{ I_{A,i}d_{A,i} \EYC(X_i;\gamma) + \frac{I_{B,i} (1-T_i)}{\pi_{B}(X_i;\alpha)\big(1-\pi_{T}(X_i;\tau)\big)} \big(Y_i-\EYC(X_i;\gamma) \big)   \Big\}$.

\subsection{Variance estimator for penalized estimating equations}

Let $N^{-1}\sum_{i=1}^N \phi(Z_i;\theta,\omega)$ be the estimation function for $\theta$, the average treatment effect, and let $N^{-1}\sum_{i=1}^N U(Z_i; \omega) + q_{\lambda}(|\omega|)sgn(\omega)$ be the penalized estimating function for $\omega$, the nuisance parameters. Since the penalty function is non-differentiable when coefficients are zero, we use the local quadratic approximation technique for the penalty function following \cite{fan2001variable} to facilitate the calculation of the differentiation. If $\omega \neq 0$ and $\omega$ is close to $\omega^\ast$, we have
\begin{gather*}
    q_{\lambda}(|\omega|)sgn(\omega) \approx \frac{q_{\lambda}(|\omega^\ast|)}{|\omega^\ast|}\omega \approx \frac{q_{\lambda}(|\omega^\ast|)}{\epsilon+ |\omega^\ast|}\omega, \\
    [q_{\lambda}(|\omega|)sgn(\omega)]' \approx \frac{q_{\lambda}(|\omega^\ast|)}{\epsilon+ |\omega^\ast|}.
\end{gather*}

For a small $\epsilon>0$, the MM-algorithm suggests that the solution $\hat{\omega}$ for the penalized estimating equations satisfies the following condition \cite{wang2012penalized,yang2020doubly}:
\begin{equation*}
    0 = N^{-1}\sum_{i=1}^N U(Z_i; \hat{\omega}) + \frac{q_{\lambda}(|\hat{\omega}|)}{\epsilon+|\hat{\omega}|}\hat{\omega} = N^{-1}\sum_{i=1}^N U^p (Z_i; \hat{\omega}).
\end{equation*}
Therefore,
\begin{equation*}
    0 = N^{-1/2}\sum_{i=1}^N U^p(Z_i; \omega^\ast) + N^{-1/2}\sum_{i=1}^N \frac{\partial U^p(Z_i; \omega^\ast) }{ \partial \omega^{\top}} (\hat{\omega} - \omega^\ast) +o_p(1).
\end{equation*}

Moving the terms, 
\begin{equation*}
    \sqrt{N} (\hat{\omega} - \omega^\ast) = \Big\{ - N^{-1} \sum_{i=1}^N \frac{\partial U^p (Z_i;\omega^\ast)}{\partial \omega^{\top}} \Big\}^{-1} \Big\{ N^{-1/2} \sum_{i=1}^N U^p(Z_i; \omega^\ast) \Big\} + o_p(1),
\end{equation*}
\noindent where 
\begin{gather*}
    \frac{\partial U^p (Z_i;\omega^\ast)}{\partial \omega^{\top}}= \frac{\partial U (Z_i;\omega^\ast)}{\partial \omega^{\top}} + E_N(\omega^\ast), \\ 
    E_N(\omega^\ast) = diag(\frac{q_{\lambda}(|\omega_1^\ast|)}{\epsilon+ |\omega_1^\ast|},...,\frac{q_{\lambda}(|\omega_p^\ast|)}{\epsilon+ |\omega_p^\ast|}), \\ 
    U^p(Z_i; \omega^\ast) = U(Z_i; \omega^\ast) + \frac{q_{\lambda}(|{\omega^\ast}|)}{\epsilon+|\omega^\ast|}\omega^\ast = U(Z_i; \omega^\ast). 
\end{gather*}

Therefore, 
\begin{equation} \label{omega_dist}
    \sqrt{N} (\hat{\omega} - \omega^\ast) = N^{-1/2} \sum_{i=1}^N -\Big[ E\Big\{ \frac{\partial U (Z_i;\omega^\ast)}{\partial \omega^{\top}} \Big\} + E_N (\omega^\ast) \Big]^{-1} U(Z_i; \omega^\ast) + o_p(1).
\end{equation}

We now consider $\theta$. The Taylor series expansion for $N^{-1/2}\sum_{i=1}^N \phi(Z_i;\hat{\theta},\hat{\omega})$ is:
\begin{align*}\label{expansion_theta}
    0 & = N^{-1/2}\sum_{i=1}^N \phi(Z_i;\theta^\ast,\omega^\ast) + N^{-1/2}\sum_{i=1}^N \frac{\partial \phi(Z_i;\theta^\ast,\omega^\ast)}{\partial \theta^{\top}} (\hat{\theta} - \theta^\ast) + N^{-1/2}\sum_{i=1}^N \frac{\partial \phi(Z_i;\theta^\ast,\omega^\ast)}{\partial \omega^{\top}} (\hat{\omega} - \omega^\ast) + o_p(1), 
\end{align*}
\noindent where $\frac{\partial \phi(Z_i;\theta^\ast,\omega^\ast)}{\partial \theta^{\top}} = 1$ (or $-1$). Therefore, using the result in equation \eqref{omega_dist}, we have:
\begin{equation*}
    \sqrt{N}(\hat{\theta} - \theta^\ast) = - N^{-1/2} \sum_{i=1}^N \Big[ \phi(Z_i;\theta^\ast,\omega^\ast) - E\big\{ \frac{\partial \phi(Z_i;\theta^\ast,\omega^\ast)}{\partial \omega^{\top}} \big\} \big\{ E\big( \frac{\partial U (Z_i;\omega^\ast)}{\partial \omega^{\top}} \big) + E_N (\omega^\ast) \big\}^{-1} U(Z_i; \omega^\ast)\Big] + o_p(1).
\end{equation*}
It follows that the variance estimator for $\hat{\theta}$ is:
\begin{equation*}
    \widehat{Var} (\hat{\theta}) = N^{-2} \sum_{i=1}^N \Big[ \phi(Z_i;\hat{\theta},\hat{\omega}) - \Big\{N^{-1}\sum_{i=1}^N \frac{\partial \phi(Z_i;\hat{\theta},\hat{\omega})}{\partial \omega^{\top}} \Big\} \Big\{ \big( N^{-1}\sum_{i=1}^N \frac{\partial U (Z_i;\hat{\omega})}{\partial \omega^{\top}} \big)
     + E_N (\hat{\omega}) \Big\}^{-1} U(Z_i; \hat{\omega})\Big]^2.
\end{equation*}

\noindent  \textit{OR estimator using the NHANES and EHR data:}
\begin{equation*}
    \sum_{i=1}^N \phi(Z_i;\hat{\theta},\hat{\omega}) = \sum_{i=1}^N \Big\{ \theta - I_{A,i}d_{A,i} \big(\EYT(X_i;\beta) - \EYC(X_i; \gamma) \big) \Big\},
\end{equation*}
where $\omega = (\beta, \gamma)$ and
\begin{align*}
    \sum_{i=1}^{N} U(Z_i; \omega) & = \sum_{i=1}^{N}
    \begin{pmatrix}
       I_{B,i}T_i\big(Y_i-\EYT(X_i;\beta)\big)X_i \\
       I_{B,i}(1-T_i)\big(Y_i-\EYC(X_i;\gamma)\big)X_i
    \end{pmatrix}.
\end{align*}

\noindent  \textit{IPW estimator using the NHANES and EHR data:}
\begin{equation*}
    \sum_{i=1}^N \phi(Z_i;\hat{\theta},\hat{\omega}) = \sum_{i=1}^N \bigg[ \theta -  \bigg\{ \frac{I_{B,i} T_i}{\pi_{B}(X_i;\alpha)\pi_{T}(X_i;\tau)}Y_i - \frac{I_{B,i}(1-T_i)}{\pi_{B}(X_i;\alpha)\big(1-\pi_{T}(X_i;\tau)\big)}Y_i \bigg\} \bigg],
\end{equation*}
where $\omega = (\alpha, \tau)$ and 
\begin{align*}
    \sum_{i=1}^{N} U(Z_i; \omega) & = \sum_{i=1}^{N}
    \begin{pmatrix}
       I_{A,i}d_{A,i}X_i - \frac{I_{B,i}T_i}{\pi_B{X_i;\alpha} \pi_T{X_i;\tau}} X_i \\
       I_{A,i}d_{A,i}X_i - \frac{I_{B,i}(1-T_i)}{\pi_B{X_i;\alpha} (1-\pi_T{X_i;\tau})} X_i
    \end{pmatrix}.
\end{align*}

\section{Additional simulation studies}

The linear and non-linear models for generating binary outcomes are:
\begin{enumerate}[label=(\alph*)]
    \item ${\rm logit}\big(P(Y_i=1|X_i)\big) = -1 + 0.5T_i + 0.5X_{i,1} + T\times X_{i,1} + 0.5X_{i,2} + 0.5X_{i,3} + 0.5X_{i,4} + 0.5X_{i,5}$;
    \item ${\rm logit}\big(P(Y_i=1|X_i)\big) = -3 + 0.5T_i + 0.5|X_{i,1}| + T_i\times |X_{i,1}| + 0.5|X_{i,2}| + 0.5|X_{i,3}| + 0.5|X_{i,4}| + 0.5|X_{i,5}|$.
\end{enumerate}

\begin{table}[H]
\centering
\caption{Sensitivity (SENS) and specificity (SPEC) of our proposed penalized estimating equation for \textbf{binary outcomes} based on 500 simulation replications. The evaluation metrics are sensitivity (SENS) and specificity (SPEC). The subscript $c$ stands for the correct specification of the outcome model ($OM$), selection model ($SM$), and treatment model ($TM$) while the subscript $m$ stands for model misspecification.}
\begin{tabular}{lrrrrrrrr}
  \hline
  & \begin{tabular}[c]{@{}c@{}} $\alpha$ \\ SENS \end{tabular} & \begin{tabular}[c]{@{}c@{}} $\alpha$ \\ SPEC \end{tabular} & \begin{tabular}[c]{@{}c@{}} $\tau$ \\ SENS \end{tabular} & \begin{tabular}[c]{@{}c@{}} $\tau$ \\ SPEC \end{tabular} & \begin{tabular}[c]{@{}c@{}} $\beta$ \\ SENS \end{tabular} & \begin{tabular}[c]{@{}c@{}} $\beta$ \\ SPEC \end{tabular} & \begin{tabular}[c]{@{}c@{}} $\gamma$ \\ SENS \end{tabular} & \begin{tabular}[c]{@{}c@{}} $\gamma$ \\ SPEC \end{tabular} \\ 
Case 1: $OM_c, SM_c, TM_c$ & 1 & 0.979 & 0.800 & 1 & 0.974 & 0.999 & 0.968 & 1 \\ 
  Case 2: $OM_c, SM_m, TM_c$ & 0.982 & 0.980 & 0.793 & 1 & 0.965 & 0.998 & 0.960 & 0.999 \\ 
  Case 3: $OM_c, SM_c, TM_m$ & 0.840 & 0.986 & 0.200 & 1 & 1 & 1 & 1 & 1 \\ 
  Case 4: $OM_c, SM_m, TM_m$ & 0.489 & 1 & 0.287 & 1 & 1 & 0.982 & 1 & 0.999 \\ 
  Case 5: $OM_m, SM_c, TM_c$ & 0.977 & 0.980 & 0.780 & 1 & 0.167 & 1 & 0.168 & 1 \\ 
  Case 6: $OM_m, SM_m, TM_c$ & 0.948 & 0.981 & 0.753 & 1 & 0.168 & 1 & 0.167 & 1 \\ 
  Case 7: $OM_m, SM_c, TM_m$ & 0.977 & 0.984 & 0.233 & 1 & 0.164 & 1 & 0.167 & 1 \\ 
  Case 8: $OM_m, SM_m, TM_m$ & 0.975 & 0.981 & 0.534 & 1 & 0.181 & 1 & 0.167 & 1 \\ 
  \hline
\end{tabular}
\label{selection_result_binary}
\end{table}

\begin{table}[H]
\footnotesize
\centering
\caption{Mean squared error (MSE) for non-null and null coefficients with \textbf{binary outcomes} using our proposed penalized estimating equation based on 500 simulation replications. MSE is evaluated when all models are correctly specified.}
\label{MSE_results_binary}
\begin{tabular}{cccccccc}
  \hline
  \begin{tabular}[c]{@{}c@{}} $\alpha$ \\ $MSE_{nonnull}$ \end{tabular} & \begin{tabular}[c]{@{}c@{}} $\alpha$ \\ $MSE_{null}$ \end{tabular} & \begin{tabular}[c]{@{}c@{}} $\tau$ \\ $MSE_{nonnull}$ \end{tabular} & \begin{tabular}[c]{@{}c@{}} $\tau$ \\ $MSE_{null}$ \end{tabular} & \begin{tabular}[c]{@{}c@{}} $\beta$ \\ $MSE_{nonnull}$ \end{tabular} & \begin{tabular}[c]{@{}c@{}} $\beta$ \\ $MSE_{null}$ \end{tabular} & \begin{tabular}[c]{@{}c@{}} $\gamma$ \\ $MSE_{nonnull}$ \end{tabular} & \begin{tabular}[c]{@{}c@{}} $\gamma$ \\ $MSE_{null}$ \end{tabular} \\ 
6.80E-01 & 2.53E-01 & 5.05E-01 & 0.00E+00 & 7.49E-02 & 4.48E-03 & 1.16E-01 & 9.14E-04 \\ 
   \hline
\end{tabular}
\end{table}

\begin{table}[H]
\centering
\caption{Coverage properties of the 95\% confidence interval for \textbf{binary outcomes} based on 500 replications: empirical coverage rate and empirical coverage rate $\pm$ $2\times$Monte Carlo standard error. The subscript $c$ stands for the correct specification of the outcome model ($OM$), selection model ($SM$), and treatment model ($TM$) while the subscript $m$ stands for model misspecification. }
\begin{tabular}{lr}
\hline
  Case 1: $OM_c, SM_c, TM_c$ & 0.960 (0.943,0.977) \\ 
  Case 2: $OM_c, SM_m, TM_c$ & 0.932 (0.910,0.954) \\ 
  Case 3: $OM_c, SM_c, TM_m$ & 0.962 (0.945,0.979) \\ 
  Case 4: $OM_c, SM_m, TM_m$ & 0.920 (0.896,0.944) \\ 
  Case 5: $OM_m, SM_c, TM_c$ & 0.940 (0.919,0.961) \\ 
  Case 6: $OM_m, SM_m, TM_c$ & 0.760 (0.723,0.797) \\ 
  Case 7: $OM_m, SM_c, TM_m$ & 0 (0,0) \\ 
  Case 8: $OM_m, SM_m, TM_m$ & 0.038 (0.021,0.055) \\ 
    \hline
\end{tabular}
\label{coverage_results_binary}
\end{table}

\begin{figure}[H]
    \centering
    \includegraphics[width = 16cm]{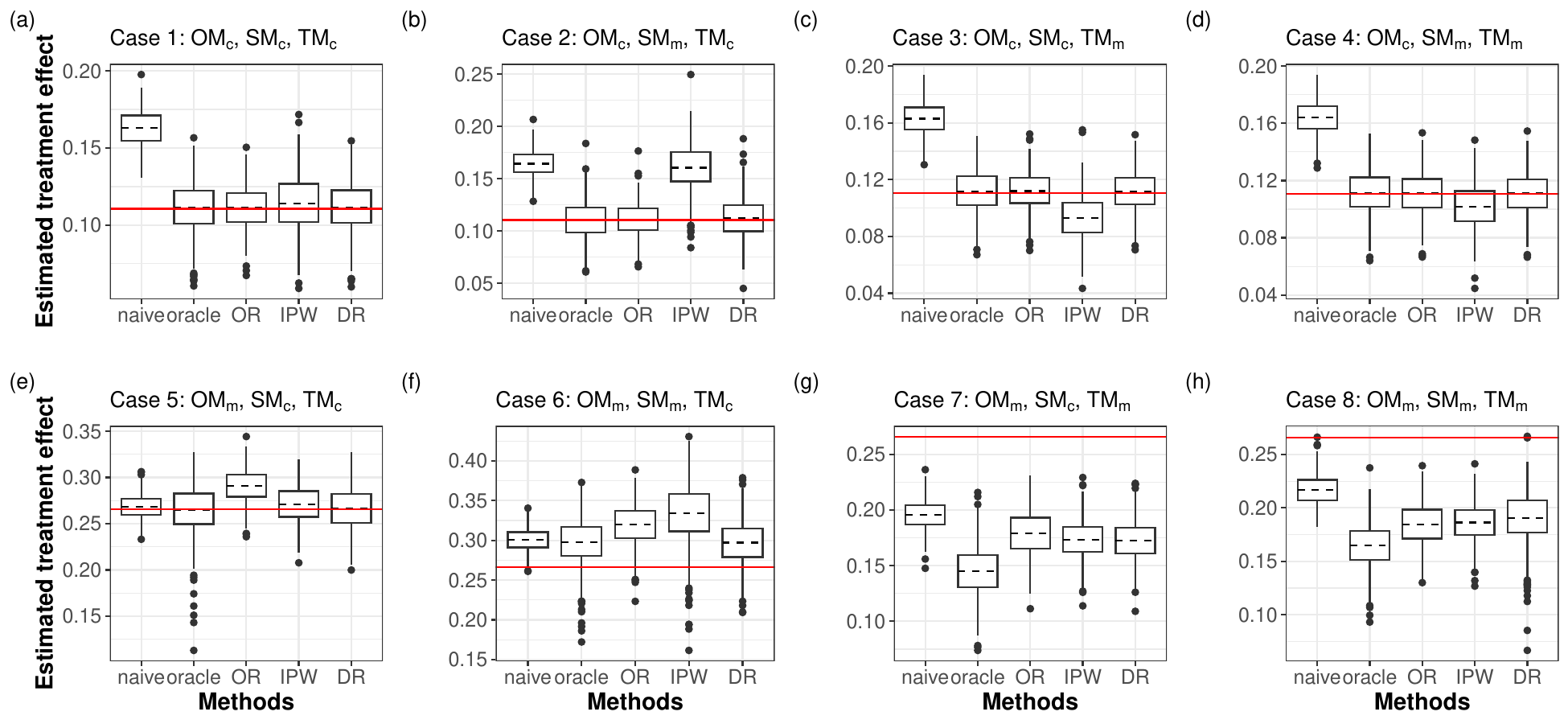}
    \caption{Estimated average treatment effect using our proposed DR estimator and four other competing estimators with \textbf{binary outcomes} based on 500 replications. The subscript $c$ stands for the correct specification of the outcome model ($OM$), selection model ($SM$), and treatment model ($TM$) while the subscript $m$ stands for model misspecification. The red horizontal line indicates the true average treatment effect.}
    \label{estimation_ATE_figure_binary}
\end{figure}

\begin{figure}[H]
    \centering
    \includegraphics[width = 16cm]{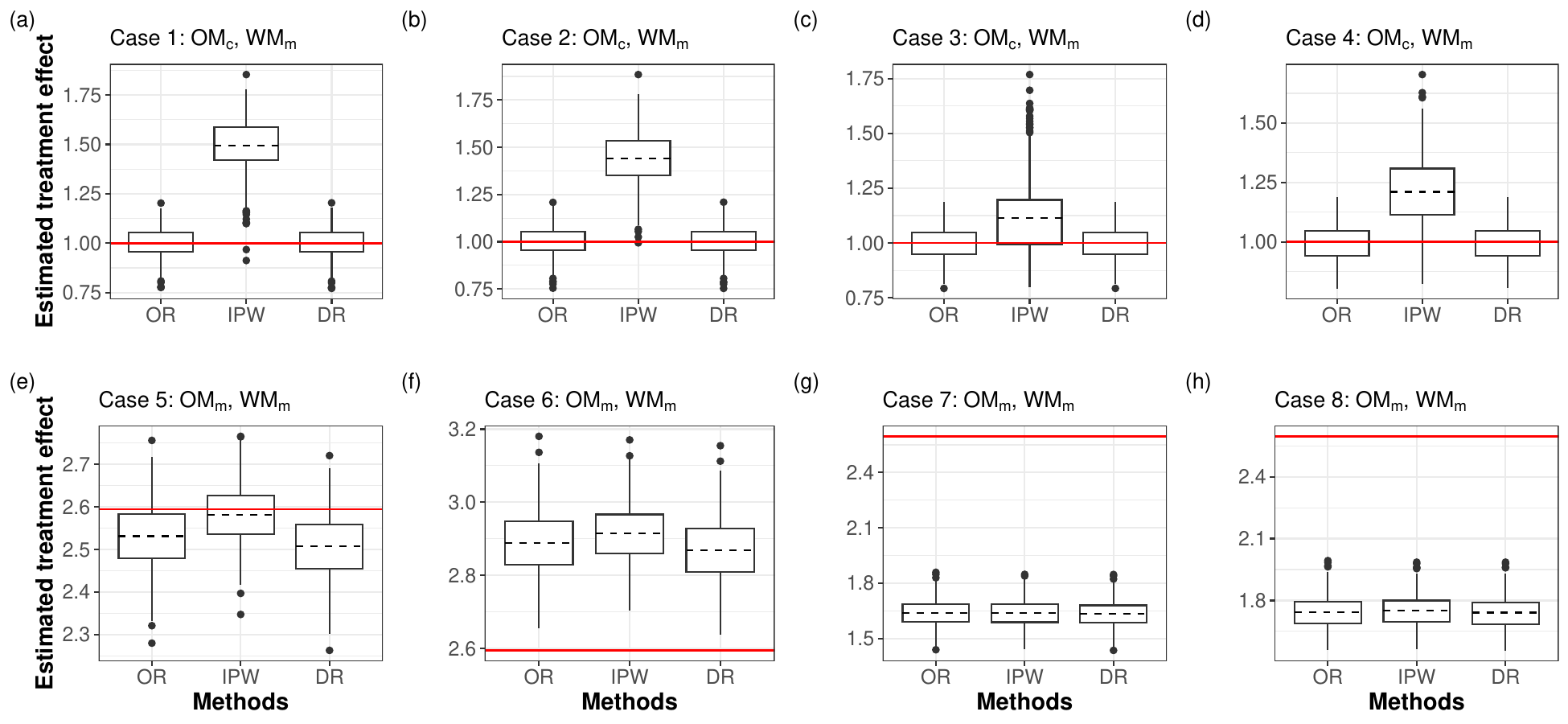}
    \caption{Estimated average treatment effect with \textbf{continuous outcomes} based on 500 replications under simulation Cases 1-8. The DR estimator is based on \textbf{joint models} of $P(I_B=1,T=t|X)$. The subscript $c$ stands for the correct specification of the outcome model ($OM$) and the weighting model ($WM$) while the subscript $m$ stands for model misspecification. The red horizontal line indicates the true average treatment effect.}
    \label{estimation_ATE_figure_binary}
\end{figure}

\begin{figure}[H]
    \centering
    \includegraphics[width = 16cm]{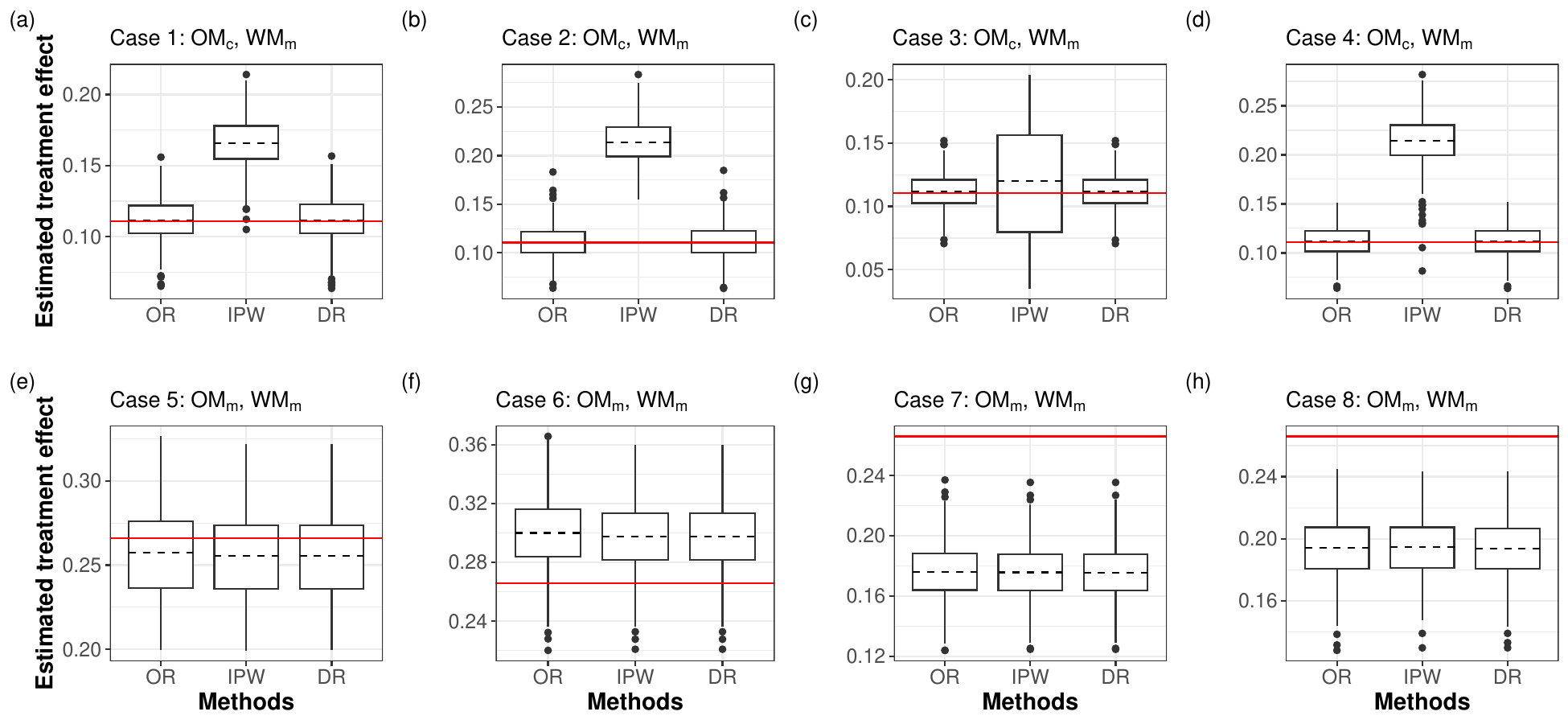}
    \caption{Estimated average treatment effect with \textbf{binary outcomes} based on 500 replications under simulation Cases 1-8. The DR estimator is based on \textbf{joint models} of $P(I_B=1,T=t|X)$. The subscript $c$ stands for the correct specification of the outcome model ($OM$) and the weighting model ($WM$) while the subscript $m$ stands for model misspecification. The red horizontal line indicates the true average treatment effect.}
    \label{estimation_ATE_figure_binary}
\end{figure}

\section{Evaluating the joint modeling approach}

\subsection{Data generation}
We first generate the complete data of the target population $\{(X_i, T_i, Y_i), i=1,...,N\}$ with $N=50,000$. More specifically, we generate $X_i = (1,X_{i,1},X_{i,2},...,X_{i,\Xlen})$ containing the intercept and $\Xlen$ covariates. Covariates, $X_{i,1},X_{i,2},...,X_{i,\Xlen}$, are generated independently from a gamma distribution with mean 0.5 and variance 0.5. The gamma distribution is chosen so that all $X_i$s are positive to facilitate generating the joint indicator of ($I_B=1,T=t$). We set $\Xlen=50$. We generate the indicator for the treated and non-treated people in sample $B$ from the following linear or non-linear weighting models ($WM$).
\begin{enumerate}[label=(\alph*)]
    \item ${\rm logit}\big(P(I_{B,i}=1, T_i=1|X_i) \big) = X_i^\top \delta_1$, where $\delta_1 = (-3.4,1,0.5,-0.5,0,...,0)^\top$; \\ 
    ${\rm logit}\big(P(I_{B,i}=1, T_i=0|X_i) \big) = X_i^\top \delta_0$, where $\delta_0 = (-2,-1,-0.5,-0.5,0,...,0)^\top$.
    \item ${\rm logit}\big(P(I_{B,i}=1, T_i=1|X_i) \big) = (X_i^\top \delta_1)^2-4.2$, where $\delta_1 = (-1,0.5,0.5,-0.5,0,...,0)^\top$; \\ 
    ${\rm logit}\big(P(I_{B,i}=1, T_i=0|X_i) \big) = (X_i^\top \delta_0)^2-4$, where $\delta_0 = (-0.5,-0.5,-0.5,0.5,0,...,0)^\top$;
\end{enumerate}

\noindent The continuous outcomes are generated from a linear or non-linear outcome model ($OM$):
\begin{enumerate}[label=(\alph*)]
    \item 
    $E(Y_i|X_i,T_i=1)= X_i^\top \beta_0$, where $\beta_0=(-0.5,1.3,0.3,0,1,1,0,...,0)^\top$; \\
    $E(Y_i|X_i,T_i=0)=X_i^\top \gamma_0$, where $\gamma_0 = (-0.5,0.3,0,0,1,1,0,..,0)^\top$.
    \item 
    $E(Y_i|X_i,T_i=1)= log\big( (X_i^\top \beta_0)^2\big)$, where $\beta_0$ is the same as (a); \\
    $E(Y_i|X_i,T_i=0)=log\big( (X_i^\top \gamma_0)^2\big)$, where $\gamma_0$ is the same as (a).
\end{enumerate}
After simulating the complete data for the target population, we draw a probability sample $A$, where the probability for each individual in the target population to be selected into sample $A$ is 0.02. The expected sample size for sample $A$, $n_A$, is 1,000. The sample size for sample $B$ is about 5,500. We have four different combinations of models for data generation (Cases S1-S4). When applying our proposed method, we always specify a linear form of covariates so that models are misspecified if the underlying true model is non-linear. We repeat each simulation scenario 500 times and the results are averaged over 500 replications.

We also perform simulation studies for binary outcomes, following the above data generation procedure except for the outcome model. The linear and non-linear models for generating binary outcomes are:
\begin{enumerate}[label=(\alph*)]
    \item 
    ${\rm logit}(Y_i|X_i,T_i=1)=X_i^\top \beta_0$, where $\beta_0=(-1.5,0.5,0.5,0.5,0.5,0,...,0)^\top$ ; \\
    ${\rm logit}(Y_i|X_i,T_i=0)=X_i^\top \gamma_0$, where $\gamma_0=(-2,0.3,0.3,0.5,0.5,0,...,0)^\top$.
    \item 
    ${\rm logit}(Y_i|X_i,T_i=1)=(X_i^\top \beta_0)^2$, where $\beta_0$ is the same as (a); \\
    ${\rm logit}(Y_i|X_i,T_i=0)=(X_i^\top\gamma_0)^2$, where $\gamma_0$ is the same as (a).
\end{enumerate}

\subsection{Simulation results}

\begin{table}[H]
\centering
\caption{Sensitivity (SENS) and specificity (SPEC) of our proposed penalized estimating equation based on \textbf{joint models} of $P(I_B=1,T=t|X)$ for \textbf{continuous outcomes}. The subscript $c$ stands for the correct specification of the outcome model ($OM$) and the weighting model ($WM$) while the subscript $m$ stands for model misspecification.}
\begin{tabular}{lrrrrrrrr}
  \hline
  & \begin{tabular}[c]{@{}c@{}} $\delta_1$ \\ SENS \end{tabular} & \begin{tabular}[c]{@{}c@{}} $\delta_1$ \\ SPEC \end{tabular} & \begin{tabular}[c]{@{}c@{}} $\tau$ \\ SENS \end{tabular} & \begin{tabular}[c]{@{}c@{}} $\tau$ \\ SPEC \end{tabular} & \begin{tabular}[c]{@{}c@{}} $\beta$ \\ SENS \end{tabular} & \begin{tabular}[c]{@{}c@{}} $\beta$ \\ SPEC \end{tabular} & \begin{tabular}[c]{@{}c@{}} $\gamma$ \\ SENS \end{tabular} & \begin{tabular}[c]{@{}c@{}} $\gamma$ \\ SPEC \end{tabular} \\ 
Case S1: $OM_c, WM_c$ & 0.981 & 0.993 & 0.958 & 0.991 & 0.993 & 0.966 & 0.978 & 0.953 \\ 
  Case S2: $OM_c, WM_m$ & 0.564 & 1 & 0.817 & 0.999 & 1 & 0.999 & 1 & 0.991 \\ 
  Case S3: $OM_m, WM_c$ & 0.982 & 0.993 & 0.959 & 0.991 & 0.998 & 0.986 & 0.99 & 0.975 \\ 
  Case S4: $OM_m, WM_m$ & 0.584 & 1 & 0.851 & 0.999 & 1 & 0.998 & 0.999 & 0.996 \\ 
  \hline
\end{tabular}
\label{setupJoint_cont_useJoint_sens}
\end{table}

\begin{table}[H]
\centering
\caption{Mean squared error (MSE) for non-null and null coefficients with \textbf{continuous outcomes} using our proposed penalized estimating equation based on \textbf{joint models} of $P(I_B=1,T=t|X)$. MSE is evaluated when all models are correctly specified.}
\label{setupJoint_cont_useJoint_MSE}
\begin{tabular}{cccccccc}
  \hline
  \begin{tabular}[c]{@{}c@{}} $\delta_1$ \\ $MSE_{nonnull}$ \end{tabular} & \begin{tabular}[c]{@{}c@{}} $\delta_1$ \\ $MSE_{null}$ \end{tabular} & \begin{tabular}[c]{@{}c@{}} $\delta_0$ \\ $MSE_{nonnull}$ \end{tabular} & \begin{tabular}[c]{@{}c@{}} $\delta_0$ \\ $MSE_{null}$ \end{tabular} & \begin{tabular}[c]{@{}c@{}} $\beta$ \\ $MSE_{nonnull}$ \end{tabular} & \begin{tabular}[c]{@{}c@{}} $\beta$ \\ $MSE_{null}$ \end{tabular} & \begin{tabular}[c]{@{}c@{}} $\gamma$ \\ $MSE_{nonnull}$ \end{tabular} & \begin{tabular}[c]{@{}c@{}} $\gamma$ \\ $MSE_{null}$ \end{tabular} \\ 
9.10E-02 & 1.24E-03 & 1.60E-01 & 1.86E-03 & 1.32E-02 & 2.68E-03 & 2.71E-02 & 5.24E-03 \\   \hline
\end{tabular}
\end{table}

\begin{table}[H]
\centering
\caption{Coverage properties of the 95\% confidence interval based on \textbf{joint models} of $P(I_B=1,T=t|X)$ for \textbf{continuous outcomes} with 500 replications: empirical coverage rate and empirical coverage rate $\pm$ $2\times$Monte Carlo standard error. The subscript $c$ stands for the correct specification of the outcome model ($OM$) and the weighting model ($WM$) while the subscript $m$ stands for model misspecification.}
\begin{tabular}{lr}
  \hline
  Case S1: $OM_c, WM_c$ & 0.954 (0.936,0.972) \\ 
  Case S2: $OM_c, WM_m$ & 0.964 (0.948,0.980) \\ 
  Case S3: $OM_m, WM_c$ & 0.972 (0.958,0.986) \\ 
  Case S4: $OM_m, WM_m$ & 0.044 (0.026,0.062) \\ 
    \hline
\end{tabular}
\label{setupJoint_cont_useJoint_coverage}
\end{table}

\newpage
\begin{figure}[H]
    \centering
    \includegraphics[width = 16cm]{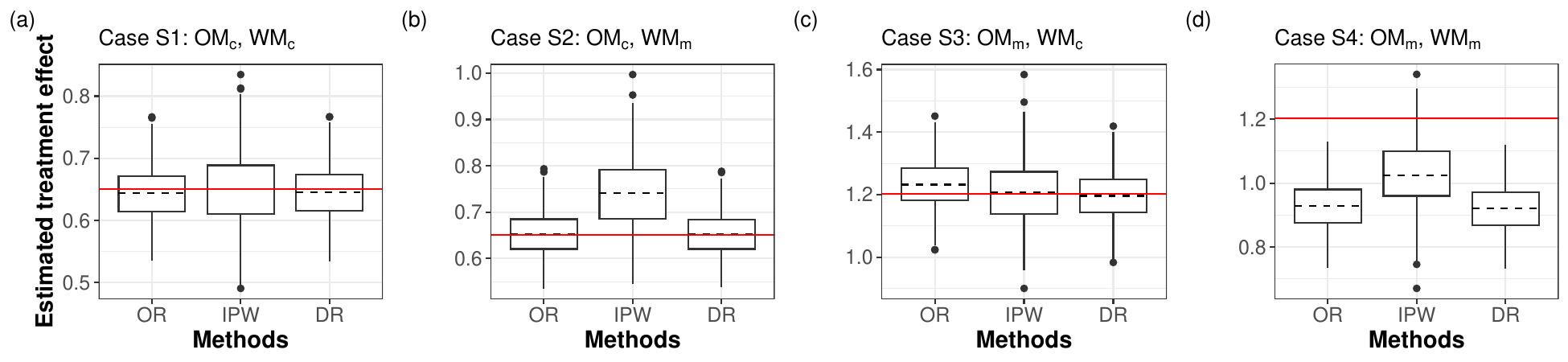}
    \caption{Estimated average treatment effect using our proposed DR estimator based on \textbf{joint models} of $P(I_B=1,T=t|X)$ and other competing estimators with \textbf{continuous outcomes} with 500 replications. The subscript $c$ stands for the correct specification of the outcome model ($OM$) and the weighting model ($WM$) while the subscript $m$ stands for model misspecification. The red horizontal line indicates the true average treatment effect.}
    \label{setupJoint_cont_useJoint_ATE}
\end{figure}

\begin{table}[H]
\centering
\caption{Sensitivity (SENS) and specificity (SPEC) of our proposed penalized estimating equation based on \textbf{joint models} of $P(I_B=1,T=t|X)$ for \textbf{binary outcomes}. The subscript $c$ stands for the correct specification of the outcome model ($OM$) and the weighting model ($WM$) while the subscript $m$ stands for model misspecification.}
\begin{tabular}{lrrrrrrrr}
  \hline
  & \begin{tabular}[c]{@{}c@{}} $\delta_1$ \\ SENS \end{tabular} & \begin{tabular}[c]{@{}c@{}} $\delta_1$ \\ SPEC \end{tabular} & \begin{tabular}[c]{@{}c@{}} $\delta_0$ \\ SENS \end{tabular} & \begin{tabular}[c]{@{}c@{}} $\delta_0$ \\ SPEC \end{tabular} & \begin{tabular}[c]{@{}c@{}} $\beta$ \\ SENS \end{tabular} & \begin{tabular}[c]{@{}c@{}} $\beta$ \\ SPEC \end{tabular} & \begin{tabular}[c]{@{}c@{}} $\gamma$ \\ SENS \end{tabular} & \begin{tabular}[c]{@{}c@{}} $\gamma$ \\ SPEC \end{tabular} \\ 
Case S1: $OM_c, WM_c$ & 0.936 & 0.988 & 0.906 & 0.982 & 0.967 & 0.959 & 0.824 & 0.923 \\ 
Case S2: $OM_c, WM_m$ & 0.500 & 1 & 0.750 & 1 & 0.999 & 0.964 & 0.956 & 0.956 \\ 
Case S3: $OM_m, WM_c$ & 0.951 & 0.986 & 0.914 & 0.982 & 0.201 & 1 & 0.674 & 0.993 \\ 
Case S4: $OM_m, WM_m$ & 0.500 & 1 & 0.750 & 1 & 0.200 & 1 & 0.603 & 1 \\ 
  \hline
\end{tabular}
\label{setupJoint_binary_useJoint_sens}
\end{table}

\begin{table}[H]
\centering
\caption{Mean squared error (MSE) for non-null and null coefficients with \textbf{binary outcomes} using our proposed penalized estimating equation based on \textbf{joint models} of $P(I_B=1,T=t|X)$. MSE is evaluated when all models are correctly specified.}
\label{setupJoint_binary_useJoint_MSE}
\begin{tabular}{cccccccc}
  \hline
  \begin{tabular}[c]{@{}c@{}} $\delta_1$ \\ $MSE_{nonnull}$ \end{tabular} & \begin{tabular}[c]{@{}c@{}} $\delta_1$ \\ $MSE_{null}$ \end{tabular} & \begin{tabular}[c]{@{}c@{}} $\delta_0$ \\ $MSE_{nonnull}$ \end{tabular} & \begin{tabular}[c]{@{}c@{}} $\delta_0$ \\ $MSE_{null}$ \end{tabular} & \begin{tabular}[c]{@{}c@{}} $\beta$ \\ $MSE_{nonnull}$ \end{tabular} & \begin{tabular}[c]{@{}c@{}} $\beta$ \\ $MSE_{null}$ \end{tabular} & \begin{tabular}[c]{@{}c@{}} $\gamma$ \\ $MSE_{nonnull}$ \end{tabular} & \begin{tabular}[c]{@{}c@{}} $\gamma$ \\ $MSE_{null}$ \end{tabular} \\ 
9.80E-02 & 1.19E-02 & 1.74E-01 & 2.13E-02 & 1.10E-01 & 3.39E-02 & 2.40E-01 & 1.04E-01 \\   \hline
\end{tabular}
\end{table}

\begin{table}[H]
\centering
\caption{Coverage properties of the 95\% confidence interval based on \textbf{joint models} of $P(I_B=1,T=t|X)$ for \textbf{binary outcomes} with 500 replications: empirical coverage rate and empirical coverage rate $\pm$ $2\times$Monte Carlo standard error. The subscript $c$ stands for the correct specification of the outcome model ($OM$) and the weighting model ($WM$) while the subscript $m$ stands for model misspecification.}
\begin{tabular}{lr}
  \hline
Case S1: $OM_c, WM_c$ & 0.948 (0.929,0.967) \\ 
Case S2: $OM_c, WM_m$ & 0.952 (0.933,0.971) \\ 
Case S3: $OM_m, WM_c$ & 0.932 (0.910,0.954) \\ 
Case S4: $OM_m, WM_m$ & 0.452 (0.408,0.496) \\ 
    \hline
\end{tabular}
\label{setupJoint_binary_useJoint_coverage}
\end{table}

\begin{figure}[H]
    \centering
    \includegraphics[width = 16cm]{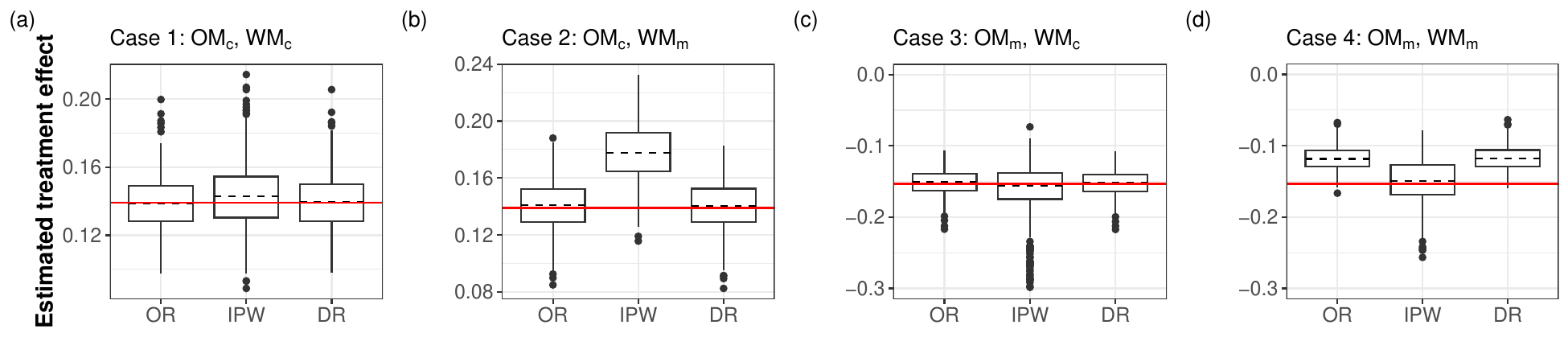}
    \caption{Estimated average treatment effect using our proposed DR estimator based on \textbf{joint models} of $P(I_B=1,T=t|X)$ and other competing estimators with \textbf{binary outcomes}. The subscript $c$ stands for the correct specification of the outcome model ($OM$) and the weighting model ($WM$) while the subscript $m$ stands for model misspecification. The red horizontal line indicates the true average treatment effect.}
\label{setupJoint_binary_useJoint_ATE}
\end{figure}

\begin{figure}[H]
    \centering
    \includegraphics[width = 16cm]{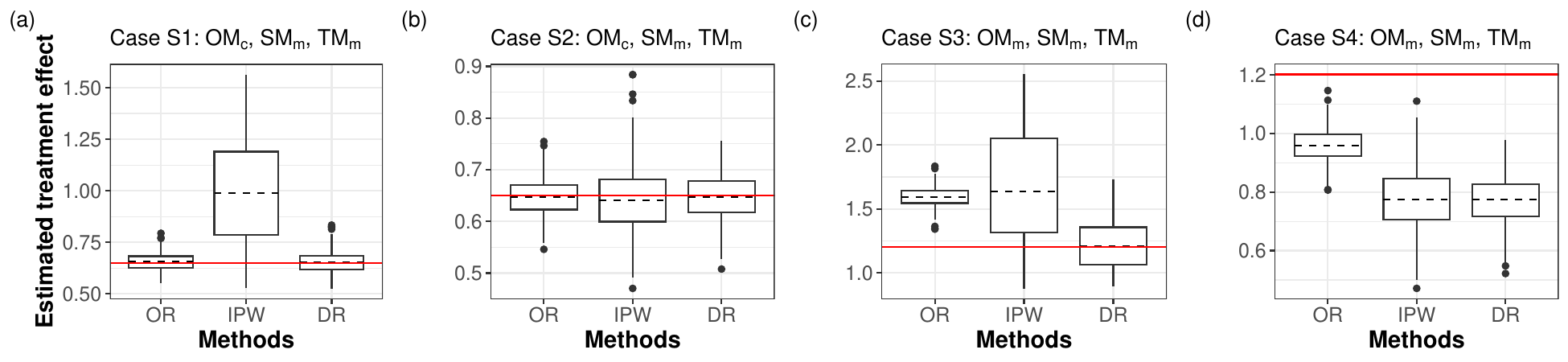}
    \caption{Estimated average treatment effect using our proposed DR estimator based on \textbf{conditional models} of $P(I_B=1|X)$ and $P(T=t|X,I_B=1)$ and other competing estimators with \textbf{continuous outcomes} under simulation Cases S1-S4. The subscript $c$ stands for the correct specification of the outcome model ($OM$) and the weighting model ($WM$) while the subscript $m$ stands for model misspecification. The red horizontal line indicates the true average treatment effect.}
\label{setupJoint_cont_useCond_ATE}
\end{figure}

\begin{figure}[H]
    \centering
    \includegraphics[width = 16cm]{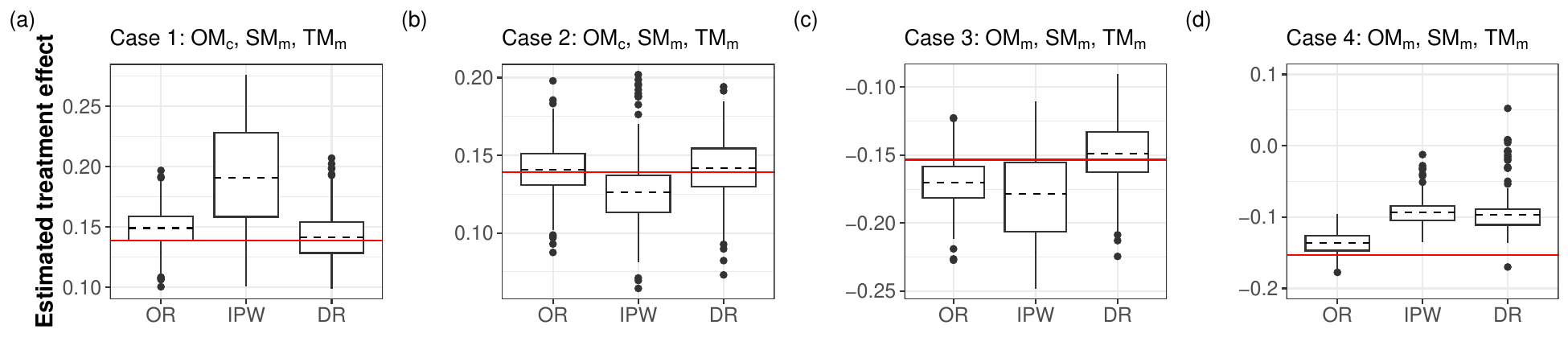}
    \caption{Estimated average treatment effect using our proposed DR estimator based on \textbf{conditional models} of $P(I_B=1|X)$ and $P(T=t|X,I_B=1)$ and other competing estimators with \textbf{binary outcomes} under simulation Cases S1-S4. The subscript $c$ stands for the correct specification of the outcome model ($OM$) and the weighting model ($WM$) while the subscript $m$ stands for model misspecification. The red horizontal line indicates the true average treatment effect.}
\label{setupJoint_binary_useCond_ATE}
\end{figure}

\section{Additional results for the data example}

\newpage

{
\begin{table}[H]
\footnotesize
\setstretch{1.5}
\caption{Estimators for the average treatment effect (ATE). }
\begin{tabular}{llll} \hline
Data source                  & \begin{tabular}[c]{@{}l@{}}Estimation of\\      nuisance parameters\end{tabular}   & ATE estimator & Expression$^{\dagger}$ \\ \hline
 \multirow{3}{*}{NHANES}      & \multirow{3}{*}{no penalty} & OR        &      $ \begin{aligned} N^{-1}\sum\limits_{i=1}^N I_{A,i}d_{A,i} \big\{ \Tilde{\EYT}(X_i^{\top}\hat{\beta}) - \Tilde{\EYC}(X_i^{\top} \hat{\gamma}) \big\}     
 \end{aligned} $      \\
 &        & IPW       &         
 $\begin{aligned}
     N^{-1} \sum\limits_{i=1}^N I_{A,i}d_{A,i} \Big\{ \frac{T_i}{ \Tilde{\pi}_T(X_i^{\top}\hat{\tau})}Y_i - \frac{1-T_i}{1-\Tilde{\pi}_T(X_i^{\top}\hat{\tau})}Y_i \Big\}
 \end{aligned}$
     \\
 &       & DR        &   
    $\begin{aligned}[b]
      N^{-1} \sum\limits_{i=1}^N I_{A,i}d_{A,i} \Big\{ 
      & \Tilde{\EYT}(X_i^{\top}\hat{\beta}) + \frac{T_i}{\Tilde{\pi}_T(X_i^{\top}\hat{\tau})} \big(Y_i-\Tilde{\EYT}(X_i^{\top}\hat{\beta}) \big)    - \\
      & \Tilde{\EYC}(X_i^{\top}\hat{\gamma}) - \frac{1-T_i}{1-\Tilde{\pi}_T(X_i^{\top}\hat{\tau})} \big(Y_i-\Tilde{\EYC}(X_i^{\top}\hat{\gamma}) \big)   \Big\} 
  \end{aligned}$
                \\ \hline \\
EHR  &        -                                           &  
\begin{tabular}[c]{@{}l@{}}
     Mean \\
     difference
\end{tabular}     &       $ \begin{aligned}
\big(\sum\limits_{i=1}^{n_B}T_i\big)^{-1} \big(\sum\limits_{i=1}^{n_B}T_iY_i \big) - \big\{\sum\limits_{i=1}^{n_B}(1-T_i)\big\}^{-1} \big\{\sum\limits_{i=1}^{n_B}(1-T_i)Y_i \big\} \end{aligned}$      \\ \\ \hline
\multirow{3}{*}{EHR}         & \multirow{3}{*}{no penalty} & OR$^{\ast}$        &       $ \begin{aligned}
n_B^{-1} \sum\limits_{i=1}^{n_B} \big\{\EYT(X_i^{\top}\hat{\beta}) - \EYC(X_i^{\top} \hat{\gamma}) \big\}     
\end{aligned}$     \\
 &                                                   & IPW       &       
 $\begin{aligned}
     n_B^{-1} \sum\limits_{i=1}^{n_B} \Big\{ \frac{T_i}{\pi_{T}(X_i^{\top}\hat{\tau})}Y_i - \frac{1-T_i}{1-\pi_{T}(X_i^{\top}\hat{\tau})}Y_i \Big\}
 \end{aligned}$     \\
 &                                                   & DR        &
 $\begin{aligned}
     n_B^{-1} \sum\limits_{i=1}^{n_B} \Big\{ 
     & \EYT(X_i^{\top}\hat{\beta}) + \frac{T_i}{\pi_{T}(X_i^{\top}\hat{\tau})} \big(Y_i-\EYT(X_i^{\top}\hat{\beta}) \big)   -\\
     & \EYC(X_i^{\top}\hat{\gamma}) - \frac{1-T_i}{1-\pi_{T}(X_i^{\top}\hat{\tau})} \big(Y_i-\EYC(X_i^{\top}\hat{\gamma}) \big)   \Big\}
 \end{aligned}
 $ \\ \hline
 \multirow{3}{*}{\begin{tabular}[c]{@{}l@{}} NHANES \\ \&EHR \end{tabular}} & \multirow{3}{*}{no penalty} & OR        &            $
 \begin{aligned}
 N^{-1}\sum\limits_{i=1}^N I_{A,i}d_{A,i} \big\{\EYT(X_i^{\top}\hat{\beta}) - \EYC(X_i^{\top} \hat{\gamma}) \big\}     
 \end{aligned}$ \\
  &                                                   & IPW       &       
  $\begin{aligned}
      N^{-1} \sum\limits_{i=1}^N \Big\{ \frac{I_{B,i} T_i}{\pi_{B}(X_i^{\top}\hat{\alpha})\pi_{T}(X_i^{\top}\hat{\tau})}Y_i - \frac{I_{B,i}(1-T_i)}{\pi_{B}(X_i^{\top}\hat{\alpha})\big(1-\pi_{T}(X_i^{\top}\hat{\tau}) \big)}Y_i \Big\}
  \end{aligned}$ \\
     &            & DR        &     
     $\begin{aligned}
         N^{-1} \sum\limits_{i=1}^{N} \Big\{ 
         & I_{A,i}d_{A,i} \EYT(X_i^{\top}\hat{\beta}) + \frac{I_{B,i} T_i}{\pi_{B}(X_i^{\top}\hat{\alpha})\pi_{T}(X_i^{\top}\hat{\tau})} \big(Y_i-\EYT(X_i^{\top}\hat{\beta}) \big)   -\\
         & I_{A,i}d_{A,i} \EYC(X_i^{\top}\hat{\gamma}) - \frac{I_{B,i} (1-T_i)}{\pi_{B}(X_i^{\top}\hat{\alpha})\big(1-\pi_{T}(X_i^{\top}\hat{\tau})\big)} \big(Y_i-\EYC(X_i^{\top}\hat{\gamma}) \big)   \Big\} 
     \end{aligned}$     \\ \hline
 \multirow{3}{*}{\begin{tabular}[c]{@{}l@{}} NHANES \\ \&EHR \end{tabular}} & \multirow{3}{*}{penalized}   & OR$^{\ast \ast}$        &      $N^{-1}\sum\limits_{i=1}^N I_{A,i}d_{A,i} \big\{\EYT(X_i^{\top}\hat{\beta}^p) - \EYC(X_i^{\top} \hat{\gamma}^p) \big\}$      \\
  &                                                   & IPW$^{\ddagger}$       &  
  $\begin{aligned}
      N^{-1} \sum\limits_{i=1}^N \Big\{ \frac{I_{B,i} T_i}{\pi_{B}(X_i^{\top}\hat{\alpha}^p)\pi_{T}(X_i^{\top}\hat{\tau}^p)}Y_i - \frac{I_{B,i}(1-T_i)}{\pi_{B}(X_i^{\top}\hat{\alpha}^p)\big(1-\pi_{T}(X_i^{\top}\hat{\tau}^p)\big)}Y_i \Big\}
  \end{aligned}$ \\
 &             & DR        &           
     $\begin{aligned}
         N^{-1} \sum\limits_{i=1}^{N} \Big\{ 
         & I_{A,i}d_{A,i} \EYT(X_i^{\top}\hat{\beta}^p) + \frac{I_{B,i} T_i}{\pi_{B}(X_i^{\top}\hat{\alpha}^p)\pi_{T}(X_i^{\top}\hat{\tau}^p)} \big(Y_i-\EYT(X_i^{\top}\hat{\beta}^p) \big)   -\\
         & I_{A,i}d_{A,i} \EYC(X_i^{\top}\hat{\gamma}^p) - \frac{I_{B,i} (1-T_i)}{\pi_{B}(X_i^{\top}\hat{\alpha}^p)\big(1-\pi_{T}(X_i^{\top}\hat{\tau}^p)\big)} \big(Y_i-\EYC(X_i^{\top}\hat{\gamma}^p) \big)   \Big\} 
     \end{aligned}$ \\ \hline
\end{tabular}
\label{estimator_list}\\
{$\dagger$: $\Tilde{\EYT}$ and $\Tilde{\EYC}$ are the outcome model for the treated and control group in the NHANES data, respectively.  $\Tilde{\pi}_T$ is the treatment model in the NHANES data. \\
$\ast$, $\ast \ast$, $\ddagger$: corresponding to the naive estimator, the OR estimator, and the IPW estimator in the simulation section.}
\end{table}
}

\newpage
\begin{figure}[H]
    \centering
    \includegraphics[width = 16cm]{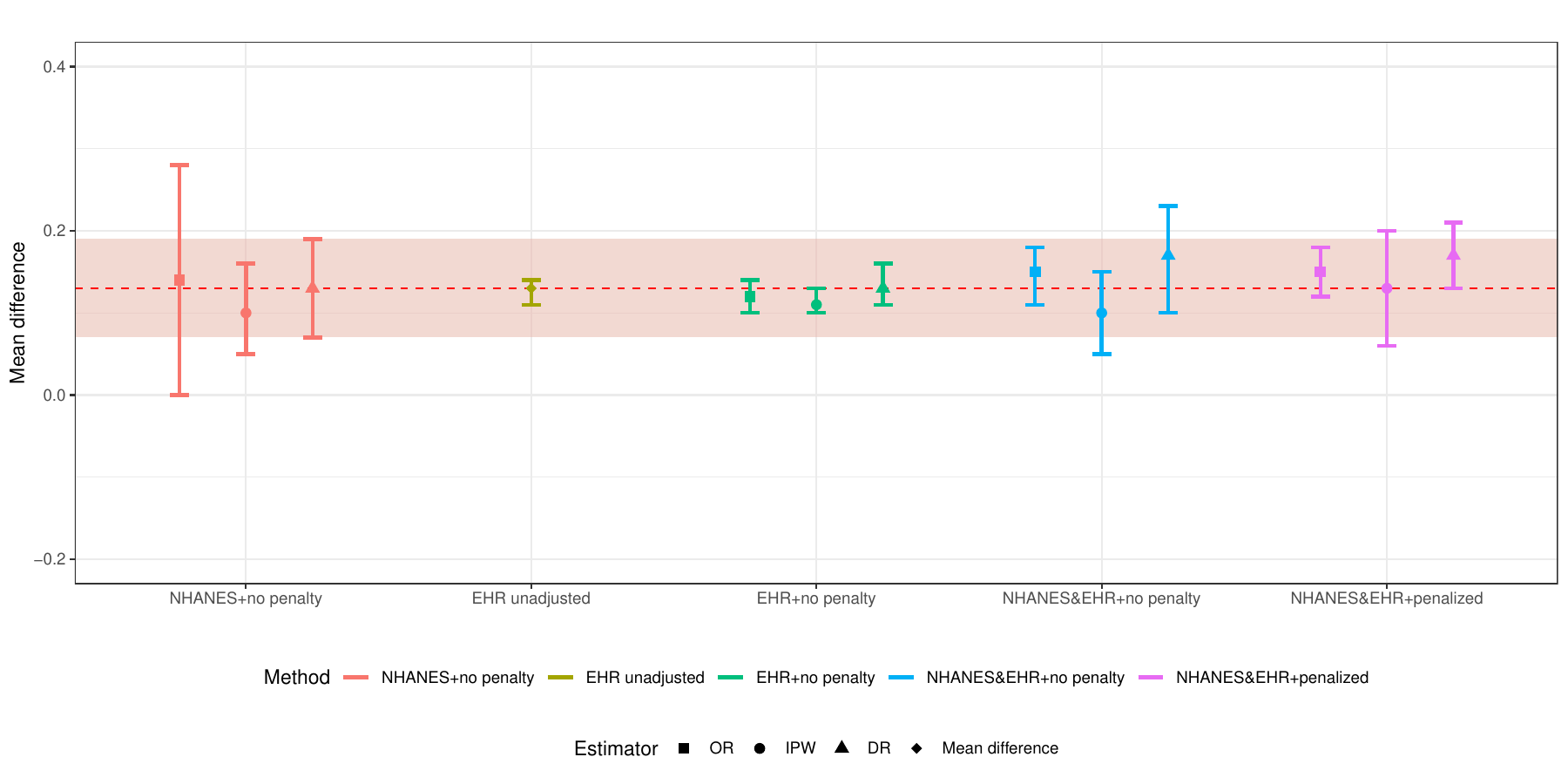}
    \caption{The estimated average treatment effect of severe obesity on hypertension and 95\% confidence interval (CI). The red horizontal line denotes the point estimate from the doubly robust estimator using exclusively the NHANES data, with the 95\% CI plotted in the banded area. The color of lines represents the method groups based on the data source (NHANES, EHR data from MGI, or both) and the use of penalization (no penalty or penalized). The ``EHR unadjusted'' method represents the sample average of the EHR data that ignores both confounding and selection bias. The shape of the point estimate represents the type of the estimator: outcome regression (OR) estimator, inverse probability weighting (IPW) estimator, and doubly robust (DR) estimator.}
    \label{hypertension_ATE_plot}
\end{figure}

\begin{table}[H]
\centering
\caption{Estimates of the average treatment effect for continuous and binary outcomes via conditional models of $P(I_B=1|X)$ and $P(T=t|X,I_B=1)$ or via joint models of $P(I_B=1,T=t|X)$. }
\begin{tabular}{cr}
\hline
\multicolumn{2}{c}{\textbf{Continuous outcome}}  \\
Conditional models & 5.20 (3.87, 6.54)\\
Joint models & 5.17 (3.87, 6.48)\\
\multicolumn{2}{c}{\textbf{Binary outcome}}  \\
Conditional models & 0.17 (0.13, 0.21)\\
Joint models & 0.17 (0.13, 0.21)\\
    \hline
\end{tabular}
\label{coverage_results_binary}
\end{table}

\printbibliography